 \documentclass[10pt,twocolumn,twoside]{IEEEtran}

\usepackage{cite}

\ifCLASSINFOpdf
\else
\fi

\usepackage{pgfplots}
\usepackage[utf8]{inputenc}
\usepackage{etex}
\usepackage{booktabs}
\usepackage{amsmath,amssymb,amsfonts}
\usepackage{mathtools}
\usepackage{amsthm}
\usepackage{xcolor}
\usepackage{mathtools}
\usepackage{macros}
\usepackage{multirow}
\usepackage{multicol}
\usepackage{kantlipsum,cuted}
\usepackage{relsize}
\usepackage{mwe}
  
\everymath{ \displaystyle}
\RequirePackage{etex} 
\usepackage[ruled,vlined]{algorithm2e}
\usepackage{subcaption}

\usepackage{tikz}
\usetikzlibrary{positioning,calc}
\usepackage[T1]{fontenc}
\usepgfplotslibrary{fillbetween}
\usepackage{diagbox} 
\usepackage{balance}
\usetikzlibrary{arrows,automata,shapes,backgrounds}
\theoremstyle{remark}
\newtheorem{remark}{Remark}
\pgfplotsset{compat=1.9}
\RequirePackage{etex} 
\newcommand{\added}[1]{#1}
\title{Distributed Policy Synthesis of Multi-Agent Systems with Graph Temporal Logic Specifications} %

\begin{document}


\author{Murat~Cubuktepe,~\IEEEmembership{Student~Member,~IEEE},~Zhe~Xu,~\IEEEmembership{Member,~IEEE},~and~Ufuk~Topcu,~\IEEEmembership{Member,~IEEE}
\thanks{Murat Cubuktepe and Ufuk Topcu are with the Department of Aerospace Engineering
and Engineering Mechanics, and the Oden Institute for Computational Engineering and Sciences, University of Texas, Austin, 201 E 24th St, Austin, TX 78712.
Zhe Xu is with the School for Engineering of Matter, Transport and Energy, Arizona State University, Tempe, AZ 85287. email: {\tt\small $\{$mcubuktepe, utopcu$\}$@utexas.edu, xzhe1@asu.edu}.

This  work  was  supported in part  by  ONR N00014-19-1-2054, NSF 1646522, NSF 1652113.

Portions of this paper previously appeared as a conference paper: M. Cubuktepe, Z. Xu and U. Topcu, Policy Synthesis for Factored MDPs with Graph Temporal Logic Specifications, \textit{Proc. International Conference on Autonomous Agents and Multi-Agent Systems
} (AAMAS), Auckland, New Zealand, 2020.}}%

\maketitle


%
\begin{abstract}
We study the distributed synthesis of policies for multi-agent systems to perform \emph{spatial-temporal} tasks. 
We formalize the synthesis problem as a \emph{factored} Markov decision process subject to \emph{graph temporal logic} specifications. 
The transition function and task of each agent are functions of the agent itself and its neighboring agents. 
In this work, we develop another distributed synthesis method, which improves the scalability and runtime by two orders of magnitude compared to our prior work.
The synthesis method decomposes the problem into a set
of smaller problems, one for each agent by leveraging the structure in the model, and the specifications.
We show that the running time of the method is linear in the number of agents.
The size of the problem for each agent is exponential only in the number of neighboring agents, which is typically much smaller than the number of agents.
\added{We demonstrate the applicability of the method in case studies on disease control, urban security, and search and rescue. 
The numerical examples show that the method scales to hundreds of agents with hundreds of states per agent and can also handle significantly larger state spaces than our prior work.}

\end{abstract}


%

\section{Introduction}

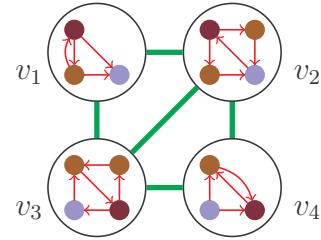
\begin{figure}[t]
	\centering
	\begin{tikzpicture}[innerblue/.style={minimum size=0.6cm,circle,draw=blue!40,fill=blue!40,thick,scale=0.4},inner/.style={minimum size=0.6cm,circle,draw=blue!50,fill=blue!50,thick,scale=0.4},innergreen/.style={minimum size=0.6cm,circle,draw=purple!50!black,fill=purple!50!black,thick,scale=0.4},innerred/.style={minimum size=0.6cm,circle,draw=orange!60!black,fill=orange!60!black,thick,scale=0.4},innerred1/.style={minimum size=0.6cm,circle,draw=red!20,fill=red!20,thick,scale=0.4}]
  \tikzstyle{every state}=[minimum size=1.3cm, fill=none,node distance=1.8cm,semithick,font=\large]
  \tikzstyle{every node}=[fill=none,semithick,font=\large]
\node[innergreen] (s11) at (-0.3,0.3) {};
\node[innerred] (s12) at (-0.3,-0.3) {};
\node[innerblue] (s13) at (0.3,-0.3) {};
\draw[->,semithick,red] (s11) -- (s12) {};
\draw[->,semithick,red,bend left] (s12) to (s11) {};
\draw[->,semithick,red] (s11) -- (s13) {};
\draw[->,semithick,red] (s12) -- (s13) {};

\node[innergreen] (s21) at (1.5,0.3) {};
\node[innerred] (s22) at (1.5,-0.3) {};
\node[innerblue] (s23) at (2.1,-0.3) {};
\node[innerred] (s24) at (2.1,0.3) {};

\draw[->,semithick,red] (s21) -- (s22) {};
\draw[->,semithick,red] (s23) -- (s21) {};
\draw[->,semithick,red] (s22) -- (s23) {};
\draw[->,semithick,red] (s21) -- (s24) {};
\draw[->,semithick,red] (s24) -- (s23) {};

\node[innerred] (s31) at (-0.3,-1.5) {};
\node[innerblue] (s32) at (-0.3,-2.1) {};
\node[innergreen] (s33) at (0.3,-2.1) {};
\node[innerred] (s34) at (0.3,-1.5) {};

\draw[->,semithick,red] (s32) -- (s31) {};
\draw[->,semithick,red] (s31) -- (s33) {};
\draw[->,semithick,red] (s33) -- (s32) {};
\draw[->,semithick,red] (s34) -- (s31) {};
\draw[->,semithick,red] (s33) -- (s34) {};

\node[innerred] (s41) at (1.5,-1.5) {};
\node[innerblue] (s42) at (1.5,-2.1) {};
\node[innergreen] (s43) at (2.1,-2.1) {};
\draw[->,semithick,red] (s42) -- (s41) {};
\draw[->,semithick,red] (s43) -- (s41) {};
\draw[->,semithick,red] (s42) -- (s43) {};
\draw[->,semithick,red,bend left] (s41) to (s43) {};

  \node[state] (1)                    {};
  \node[state]         (2) [right of=1] {}; 
  \node[state]         (3) [below of=1] {}; 
  \node[state]         (4) [right of=3] {};;
  \draw[line width=2pt,green] (1) -- (2); 
  \draw[line width=2pt,green] (2) -- (4);
  \draw[line width=2pt,green] (2) -- (3);
  \draw[line width=2pt,green] (3) -- (1);
  \draw[line width=2pt,green] (4) -- (3);
  \node [draw=none] at (0.8,-0.5) {${}$};
  \node [draw=none] at (2.6,0.2) {${}$};
  \node [draw=none] at (3.4,-0.6) {${}$};
  \node [draw=none] at (-0.9,-0.3) {$v_1$};
  \node [draw=none] at (2.8,-0.3) {$v_2$};
  \node [draw=none] at (-0.9,-2.1) {$v_3$};
  \node [draw=none] at (2.8,-2.1) {$v_4$};
  \node [draw=none] at (-1.35,0.3) {};
  \node [draw=none] at (2.8,0.3) {};
  \node [draw=none] at (-0.9,-1.5) {};
  \node [draw=none] at (2.8,-1.5) {};
\end{tikzpicture}
\caption{An example of an undirected graph. The nodes ($v_1, v_2, v_3,$ and $v_4$) of the graph represents the agents. Each agent has its own set of states and actions. We draw an edge between two agents if they share the current state information and task. The color of the states gives the label for each agent.} 
	\label{graph_exp}
\end{figure}

\added{We study the distributed policy synthesis of multi-agent systems where multiple agents coordinate to achieve a set of \emph{spatial-temporal tasks}.}
We define the spatial-temporal tasks over an underlying graph modeling the interaction between the agents.
For example, in Fig.~\ref{graph_exp}, each node of the graph represents an agent, and each agent has its own set of states and actions. 
We draw edges between \emph{neighboring} agents that exchange their current state information and share a task. 
For each agent, the transition function between its states may depend on its current state and the neighboring agents.
The labels at each node and edge of the graph provide information about the task of each agent.
The spatial-temporal task for each agent depends both on the agent itself and its neighboring agents.
\added{The different nodes in a graph in Fig.~\ref{graph_exp} can model different police officers.}
The states of each node represents the intersections that the corresponding police officer is monitoring in a city.
An edge between two nodes exists if the two corresponding police officers can share their state information and tasks.
An example of a spatial-temporal task is \textquotedblleft agent $v_3$, or a neighboring agent of $v_3$ should be in the brown intersection in every three time steps\textquotedblright.


We model each agent's behavior as a Markov decision process (MDP)~\cite{puterman2014markov}, a widely used model to solve decision-making problems in dynamic environments.
A possible way to represent the states of the overall multi-agent system is enumerating all possible states of the agents.
However, the size of the composed MDP will scale exponentially in the number of agents, and the representation will be impractical for policy synthesis.
In many systems, the interaction between the agents is typically \emph{sparse}, meaning each agent in the multi-agent system shares a task and exchanges their current state information with only a few other agents. 
Examples of such systems appear in biochemical networks~\cite{prescott2014layered}, smart grids~\cite{dorfler2014sparsity}, swarm robots~\cite{pickem2017robotarium}, and disease control~\cite{choisy2007probabilistic}. 

By exploiting the sparsity in the interaction between agents, we represent the composed MDP as a factored MDP~\cite{guestrin2003efficient,cheng2013variational,guestrin2002multiagent,forsell2006approximate}. 
\added{According to~\cite{guestrin2003efficient}, the state-space and the transition model of a factored MDP can be represented as a dynamic Bayesian network~\cite{dean1989model}, which is a type of probabilistic graphical model.
We also utilize the structure of the underlying network to represent the spatial-temporal tasks as a function of the neighboring agents, which allows us to develop algorithms that scale to a large number of agents.
These representations alleviate the need to enumerate all possible states of the agents and allows the synthesis of policies for systems with large groups of agents.
Similar to the approaches for solving factored MDPs, we exploit the fact that the transition dynamics of an agent depend on the neighboring agents, which is significantly smaller than the number of overall agents.
We also utilize the structure of the underlying network to represent the spatial-temporal tasks as a function of the neighboring agents, which allows us to develop algorithms that scale to a large number of agents.
These representations alleviate the need to enumerate all possible states of the agents and allows the synthesis of policies for systems with large groups of agents.}
\looseness=-1


Related work in factored MDPs focuses on computing a policy to maximize the expected reward for a given reward function.
A wide range of tasks, such as avoiding certain parts of an environment, cannot be expressed by any reward function~\cite{littman2017environment,hahn2019omega}.
Linear temporal logic (LTL) can concisely express such tasks~\cite{Pnueli,BK08}. 
However, LTL cannot express spatial-temporal tasks involving multiple agents without an exponential blow-up in the size of the automaton.

We represent spatial-temporal tasks in a novel specification language called graph temporal logic (GTL)~\cite{zhe_GTL}. 
GTL formulas can represent tasks such as \textquotedblleft the police officer at node $v_3$ or their neighboring officers should visit intersection labeled as brown in every three hours\textquotedblright.
GTL is an extension of LTL and defines the spatial-temporal tasks over the labels on the graph. 
We use GTL formulas to express tasks that concern a number of agents on the graph.

\added{In theory, it may be possible to convert a short GTL formula to an overly lengthy LTL formula by enumerating all nodes and edges that satisfy the node and edge propositions. 
However, GTL is a logic that is specifically defined for expressing spatial-temporal tasks with a graph structure. 
Therefore, it is easier for practitioners to express such graph-temporal specifications concisely in GTL than other logics such as LTL.
Additionally, GTL can concisely define the tasks that a sub-group of the overall agents should carry out instead of all agents.}

Expressing the spatial-temporal tasks as GTL specifications gives us the following advantages. 1) GTL provides a concise way to express the spatial-temporal tasks that constrain the behavior of the agents to satisfy the spatial-temporal tasks. Recall that such tasks cannot be expressed by designing a reward function in general. 2) GTL resembles natural languages, and requirements specified by practitioners can be translated into GTL specifications. \added{3) GTL specifications can be converted into a deterministic finite automaton~\cite{zhe_GTL}, which allows us to utilize the existing techniques for policy synthesis.}


\added{In our conference paper, we synthesized a policy for each agent as a function of its current state and neighboring agents states~\cite{cubuktepe2020policy}.
However, the number of policy variables in that formulation scales exponentially both in the number of states and actions of the agent and its neighboring agents.
Moreover, the number of constraints due to the underlying MDP also scales exponentially, similar to the policy variables.
Therefore, computing policies for each agent that is a function of the states of itself and neighboring agents can be computationally challenging.
}

\added{To alleviate the resulting computational complexity, we develop a new centralized algorithm to synthesize a policy for each agent as a function of only their local states, assuming that the transition function of each agent is a function of its current state.
This formulation allows us to scale the policy synthesis to state spaces that are significantly larger than the method in the prior conference paper.
We also develop a new distributed algorithm by decomposing the centralized problem into smaller synthesis problems, one for each agent, similar to the prior approach.}

\added{We compare the runtime of the algorithms, theoretically and empirically, showing that the method requires exponentially fewer variables and constraints than prior work.
The numerical examples show that the method in this paper outperforms the method in our prior work by at least two orders of magnitude in the runtime.
We also compare the methods against approaches based on metric temporal logic (MTL), where the optimization problem is formulated as a mixed-integer linear problem. 
We show that the GTL-based formulation significantly outperforms the MTL-based formulation on problems with a large number of agents.}
\looseness=-1

We demonstrate the effectiveness of the algorithm on three examples with a large number of agents. 
In the first example, we consider disease management in crop fields with GTL specifications~\cite{sabbadin2012framework}.
A contaminated crop field can infect its neighbors, and the yield of that crop field decreases.
The GTL specifications ensure that some of the crop fields are treated immediately to prevent the spreading of disease.
We maximize the expected yield of the crop fields while treating the critical crop fields appropriately.
We show that the running time of the decentralized algorithm scales linearly with the number of agents.
In the second example, we consider an urban security problem. 
The problem is to assign patrol tasks to police officers such that an officer sufficiently monitors critical locations in the city.
We express the task of monitoring the critical locations in GTL specifications.
Finally, we consider a search and rescue scenario where a large groups of robots coordinate with each other to search victims in a Gazebo environment. 
The results show that the proposed distributed algorithm scales to hundreds of agents while ensuring that the agents achieve tasks expressed in GTL formulas.
\added{In our numerical examples, we also show that the method we develop in this paper can handle much larger state spaces than our previous result.
Specifically, the method we developed in the prior conference paper runs out of memory on the search and rescue scenario.
On the other hand, we could compute optimal policies for this scenario in minutes using the proposed method in this paper.}
\looseness=-1

\noindent\textbf{Related work.} 
Related work on factored MDPs considers maximizing the expected reward of the overall multi-agent system. 
The existing works maximize the expected reward using approximate linear
programming~\cite{guestrin2003efficient,guestrin2002multiagent,forsell2006approximate}, approximate policy iteration~\cite{sabbadin2012framework,peyrard2006mean}, and approximate value iteration~\cite{guestrin2003efficient,cheng2013variational}.
However, in general, maximizing the expected reward is not sufficient to implement spatial-temporal tasks that include multiple agents. 
\cite{hahn2019omega} shows that no reward structure can capture tasks that are given by temporal logic specifications.  
\looseness=-1

The work in~\cite{sahin2017provably} considers the problem of coordinating multiple agents subject to constraints on the number of agents achieving tasks given in temporal logic specifications. 
However, they require agents to be homogeneous and consider all agents for different tasks. 
Recent work in~\cite{haghighi2016robotic,liu2018distributed,liu2017distributed} proposes spatial-temporal logic for swarm robots. 
The proposed logic is less expressive than GTL, and they solve a mixed-integer linear program, which is significantly more challenging--in theory, and in practice--than the optimization problems in our proposed approach.
References~\cite{CensusSTL2016,swarmSTL2019} propose signal temporal logic inference methods for swarm systems, but they do not control the agents.

Reference~\cite{verginis2018motion} proposes a framework for potential-based collision avoidance of multi-agent systems with temporal logic specifications. 
The work in \cite{Allerton2019} considers a synthesis problem of a multi-agent system with temporal logic specifications, where the agents do not communicate with each other all the time.
However, both of the algorithms are centralized, and the algorithm may not scale to a large number of agents.


 
\section{Preliminaries}%
We denote a \emph{probability distribution} over a set $\distDom$ by a function $\distFunc\colon\distDom\rightarrow\Ireal$ with $\sum\nolimits_{\distDomElem\in\distDom}\distFunc(\distDomElem)=\distFunc(\distDom)=1$. 
$\Distr(\distDom)$ denotes the set of of all distributions in $\distDom$. Let $\mathbb{T}=\{1, 2, \dots\}$ be a discrete set of time indices. For two sets $A, B$ we define $A \subseteq B$ if $B$ contains all elements of $A$.%
\begin{definition}[Undirected graph]
We denote $G=(V, E)$ as an undirected graph, where $V$ is a finite set of nodes that represents the agents and $E$ is a finite set of edges. We use $e=\{v_1, v_2\}$ to denote that the edge $e\in E$ connects $v_1$ and $v_2\in V$. For agents $v_i, v_j$, we call $v_j$ a neighboring agent of $v_i$, if there is an edge $e$ that connects $v_i$ and $v_j$. Let $N(i)\subseteq V$ be the set of agent $v_i$ and the neighboring agents of the agent $v_i$. We denote $N(i,j)=N(i)\cap N(j)$ and $N(i\setminus j)=N(i)\setminus N(j)$ and $\hat{N}(i)=N(i)\setminus i$. The number of agents in $V$ is $M$. 
\end{definition}%
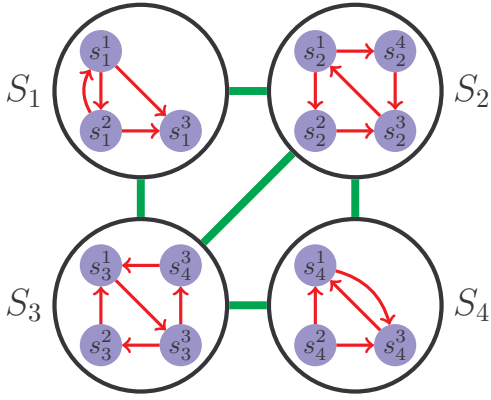
\begin{figure}[t]
    \centering%
\begin{tikzpicture}[remember picture,
  inner/.style={minimum size=0.10cm,circle,draw=blue!40,fill=blue!40,ultra thick,inner sep=0.0pt},
  outer/.style={minimum size=0.10cm,draw=black,fill=white!20,ultra thick,inner sep=0.0pt},
  blank/.style={draw=none,fill=none,ultra thick,inner sep=0.001pt}
  ]
  \node[circle,outer,draw=black] (A) { 
    \begin{tikzpicture}
      \node [inner] (ai)  {$s^1_1$};
      \node [inner,below=0.5 cm of ai] (aii) {$s^2_1$};
      \node [inner,right=0.5 cm of aii] (aiii) {$s^3_1$};
      \draw[->,red,very thick] (ai) -- (aii); 
      \draw[->,red,very thick] (ai) -- (aiii);
      \draw[->,red,very thick] (aii) -- (aiii);
      \draw[->,red,very thick,bend left] (aii) to (ai);
    \end{tikzpicture}
  };
  \node[circle,outer,draw=black,right=0.5 cm of A] (B) {
    \begin{tikzpicture}
      \node [inner] (bi)  {$s^1_2$};
      \node [inner,below=0.5 cm of bi] (bii) {$s^2_2$};
      \node [inner,right=0.5 cm of bii] (biii) {$s^3_2$};
      \node [inner,right=0.5 cm of bi] (biv) {$s^4_2$};
      \draw[->,red,very thick] (bi) -- (bii);
      \draw[->,red,very thick] (biii) -- (bi);
      \draw[->,red,very thick] (bii)-- (biii);
      \draw[->,red,very thick]  (bi) -- (biv);
      \draw[->,red,very thick]  (biv) -- (biii);    
    \end{tikzpicture}
  };
    \node[circle,outer,draw=black,below=0.5 cm of A,label={}] (C) {
    \begin{tikzpicture}
      \node [inner] (ci)  {$s^1_3$};

      \node [inner,below=0.5 cm of ci] (cii) {$s^2_3$};
      \node [inner,right=0.5 cm of cii] (ciii) {$s^3_3$};
      \node [inner,right=0.5 cm of ci] (civ) {$s^3_4$};
      \draw[->,red,very thick] (cii) -- (ci);
      \draw[->,red,very thick] (ci) -- (ciii);
      \draw[->,red,very thick] (ciii) -- (cii);
      \draw[->,red,very thick]  (civ)-- (ci);
      \draw[->,red,very thick] (ciii) -- (civ);
    \end{tikzpicture}
  };
    \node[circle,outer,draw=black,right=0.5 cm of C,label={}]  (D) {
    \begin{tikzpicture}
      \node [inner] (di)  {$s^1_4$};
      \node [inner,below=0.5 cm of di] (dii) {$s^2_4$};
      \node [inner,right=0.5 cm of dii] (diii) {$s^3_4$};
      \draw[->,red,very thick] (dii) -- (di);
      \draw[->,red,very thick] (diii) -- (di);
      \draw[->,red,very thick] (dii) -- (diii);
      \draw[->,red,very thick,bend left] (di) to (diii);
    \end{tikzpicture}
  }; 
            \node [blank,left=0.1cm of A] (AA)  {\Large $S_1$};
        \node [blank,right=0.1 of B] (BB)  {\Large $S_2$};
          \node [blank,left=0.1cm of C] (CC)  {\Large $S_3$};
        \node [blank,right=0.1 of D] (DD)  {\Large $S_4$};

   \draw[green,line width=3pt] (A) --  (B);
   \draw[green,line width=3pt] (A) --  (C);
   \draw[green,line width=3pt] (B) --  (C);
   \draw[green,line width=3pt] (B) --  (D);
   \draw[green,line width=3pt] (C) --  (D);
\end{tikzpicture}%
    \caption{An example of a factored MDP. For each agent $v_i$ in $V$, $S_i$ depicts the state space of the agent $v_i$, and $s_i^j$ depicts the state $j$ of the agent $v_i$. The arrows between the states of the agent $v_i$ shows the transitions between states of the agent $v_i$. The transition probabilities between states in $S_i$ depend on the current state of $S_i$ and the neighboring agents in the graph.}
        \label{fig:graph_mdp}
\end{figure}%
\begin{definition}[Factored MDP]
A factored Markov decision process (MDP) $\MdpInit$ on $G=(V, E)$ is given by a finite set $S$ of states, which is a Cartesian product of the states for each agent $v_i$ in $V$, i.e., $S= S_1  \times  \cdots \times S_M$, an initial state $\sinit \in S$, a finite set $\Act=\Act_1 \times \cdots \times \Act_M$ of actions, a transition function $\probmdp =\probmdp_1  \times \cdots \times \probmdp_M$, where for each agent $i$,  $\probmdp_i\colon S_{N(i)}\times\Act_{N(i)}\rightarrow\Distr(S_{N(i)})$ gives the transition function for agent $i$, where $S_{N(i)}$ and $\Act_{N(i)}$ denote the Cartesian product of the sets of states and actions of the agents in $N(i)$ respectively, a finite set $\pi$ of atomic propositions, a labeling function $\Label = \Label_1 \times \cdots \times \Label_M$, where $\Label_i \colon S_{N(i)} \rightarrow 2^{\pi}$ labels each state $s \in S_{N(i)}$ with a subset of atomic propositions $\Label_i (s) \subseteq \pi$ and a reward function $R= R_1 \times \cdots \times R_M$, where $R_i \colon S_{N(i)}\times\Act_{N(i)}\rightarrow\R_{\geq 0}$ assigns a reward to state-action pairs for agent $i$. We use $s(t)$ to denote the states of all agents in $V$ at time index $t$.
\end{definition}%
\noindent We give two examples to illustrate the concepts related to factored MDPs. The first example shows the relationship between the agents in the graph. The second example illustrates how the transition probabilities between states of an agent depend on the neighboring agents.%
\begin{example}%
    We show an example of a factored MDP in Fig.~\ref{fig:graph_mdp}. Consider a graph $G=(V,E)$, where $V=\lbrace v_1, v_2, v_3, v_4\rbrace$ and $E=\lbrace \lbrace v_1, v_2\rbrace, \lbrace v_1, v_3\rbrace, \lbrace v_2, v_4\rbrace, \lbrace v_2, v_3\rbrace, \lbrace v_3, v_4 \rbrace \rbrace.$ The black nodes in Fig.~\ref{fig:graph_mdp} represent the agents. We show the edges between agents in the graph with green lines in Fig.~\ref{fig:graph_mdp}. For $v_i \in V$, we denote the state space and action space of the agent $v_i$ as $S_i$ and $\Act_i$, and we denote the state $j$ in $S_i$ as $s^j_i$. We treat $s^j_i$ and $s^l_k$ to be different states if $i\neq k$ or $j\neq l$. For this example, $N(1)=\lbrace 1,2,3\rbrace, N(4)=\lbrace 2,3,4\rbrace$, and $N(1,4)=\lbrace 2,3\rbrace.$
    \end{example}%
\begin{example}
    We show an example of the transition function $\probmdp_i$ of the agent $v_1$ in a factored MDP in Fig.~\ref{fig:graph_mdp2}. 
    For this example, we assume that $S_{N(1)}=\lbrace  s_1,s_2,s_3\rbrace$, for $s_1 \in S_1, s_2 \in S_2, s_3 \in S_3$ and we write $\Act_{N(1)}=\Act_1$ for a special case that the transition function $\probmdp_i$ does not depend on the actions of the neighboring agents. The transition function $\probmdp_i$ is a function of the states of agent $v_1$ and its neighboring agents, and the action of agent $v_1$. 
    The transition probabilities between the states of $S_1$ are given with red lines in Fig.~\ref{fig:graph_mdp2}. 
    For example, in Fig.~\ref{fig:graph_mdp2}, the transition probabilities between states in $S_1$ for a given action $\act^1_1$ is a function of $s_1 \in S_1$, $s_2 \in S_2$ and $s_3 \in S_3$.  
\end{example}%
\begin{definition}[Policy]\label{def:scheduler} 
	A \emph{memoryless and randomized policy} for a factored MDP $\mdp$ is a function $\sched = \sched_1 \times \sched_2 \times \cdots \times \sched_M$, where for each agent i, $\sched_i \colon  S_{N(i)} \rightarrow\Distr(\Act_{N(i)})$.
The set of all policies over $\mdp$ is $\Sched_\mdp$.
\end{definition}%
Applying a policy $\sched\in\Sched_\mdp$ to a factored MDP $\mdp$ yields an \emph{induced factored Markov chain} $\mdp_\sched$. 

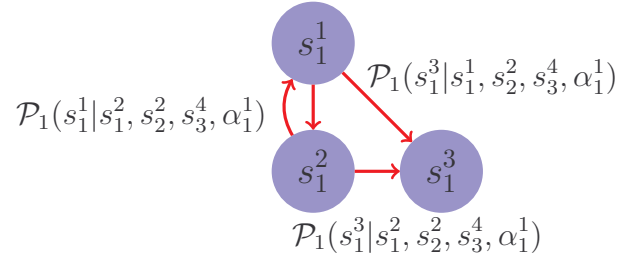
\begin{figure}[t]
    \centering
    \begin{tikzpicture}[  inner/.style={minimum size=0.17cm,circle,draw=blue!40,fill=blue!40,ultra thick,inner sep=3.5pt},  blank/.style={draw=none,fill=none,ultra thick,inner sep=0.001pt},scale=4.5]
      \node [inner] (ai)  {\Large $s^1_1$};
      \node [inner,below=0.6cm of ai] (aii) {\Large $s^2_1$};
      \node [inner,right=0.6cm of aii] (aiii) {\Large $s^3_1$};
      \draw[->,red,very thick] (ai) -- (aii); 
      \draw[->,red,very thick,bend left] (aii) to (ai); 
      \draw[->,red,very thick] (ai) -- (aiii);
      \draw[->, red,very thick] (aii) -- (aiii);
      \node[blank,below left=0.3cm and 0.2cm of ai] {\large $\probmdp_1(s^1_1|s^2_1,s^2_2,s^4_3,\act^1_1)$};
      \node[blank,below right=-0.2cm and 0.3cm of ai] {\large $\probmdp_1(s^3_1|s^1_1,s^2_2,s^4_3,\act^1_1)$};
     \node[blank,below right=0.2cm and -0.7cm of aii] {\large $\probmdp_1(s^3_1|s^2_1,s^2_2,s^4_3,\act^1_1)$};

    \end{tikzpicture}

    \caption{An example of the transitions between states for agent $v_1$ if the state is $s=2$ for agent $v_2$ and $s=4$ for agent $v_3$, and if the action $\alpha$ = 1 is taken for agent $v_1$.}
        \label{fig:graph_mdp2}
\end{figure}

\begin{definition}[Factored induced MC]\label{def:induced_dtmc}
	For a factored MDP $\MdpInit$ and a policy $\sched\in\Sched_{\mdp}$, the factored \emph{MC induced by $\mdp$ and $\sched$} is $\mdp_\sched=(S, \sinit, \Act, \probmdp_\sched, \pi, \Label, R)$, where%
	\begin{align*}
	\displaystyle	\probmdp_\sched(s' | s)=\sum\nolimits_{\act\in\Act(s)} \sched(s,\act)\cdot\probmdp(s' | s,\act) \;\forall s,s'\in S.
	\end{align*}%
\end{definition}%
\begin{definition}[Trajectory]\label{def:trajectory}
A finite or an infinite sequence $\varrho_{\sched} = s(0)s(1)s(2)\ldots$ of states in $\mathcal{M}$ that is generated by a policy $\sched$ $\in$ $\Sched_{\mathcal{M}}$ is called a \textit{trajectory}. 
\end{definition}%
For an induced factored MC $\mathcal{M}_{\sched}$, and the initial state $\sinit$, the probability of reaching state $s'$ from state $s$ at time step $t$ is equal to $\mathcal{P}_{\sched}(s'|s)$.
\subsection{Graph Temporal Logic}%
We review the theoretical framework of graph temporal logic (GTL), which was introduced in \cite{zhe_GTL}. %
 
 We denote $Y$ as the set of states for the edges, which is a Cartesian product of the states for each node $e_i$, i.e., $Y= Y_1 \times  \cdots \times Y_{|E|}$, where $Y_i$ is the state space of the edge $e_i$. 
 
\begin{definition}[Graph-temporal trajectory]
	\label{graph}
	A \textit{graph-temporal trajectory} on $G$ is a tuple $g=(s(t), y(t))$ for each time index $t \in\mathbb{T}$, where $s(t):\mathbb{T}\rightarrow S$ gives the label for each node $v_i\in V$ at each time index $t\in\mathbb{T}$, and $y(t):\mathbb{T}\rightarrow Y$ gives the label for each edge $e_i\in E$ at each time index $t\in\mathbb{T}$. 
\end{definition}%

\begin{definition}[Node and edge propositions]
	\label{node}
	An \textit{atomic node proposition} $\pi$ is a Boolean valued map on $S$, and an \textit{edge proposition} $\rho$ is a Boolean valued map on $Y$.
\end{definition}%


A graph-temporal trajectory $g=(s(t), y(t))$ satisfies $\pi$ at a node $v$ and at a time index $t$, which is denoted as $(g,v,t)\models\pi$, if and only if $s_v(t) \in \mathcal{O}(\pi)$, which denotes the subset of states $S$ for which $\pi$ is true.
Similarly, a graph-temporal trajectory $g=(s(t), y(t))$ satisfies $\rho$ at an edge $e$ and at a time index $t$, which is denoted as $(g,e,t)\models\rho$, if and only if $y_e(t) \in \mathcal{O}(\rho)$, which denotes the subset of edges $Y$ for which $\rho$ is true.%

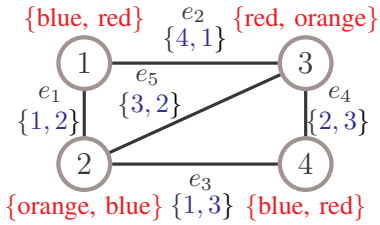
\begin{figure}[t]
    \centering

    \begin{tikzpicture}[  inner/.style={minimum size=0.17cm,circle,draw=black!40,fill=white!40,ultra thick,inner sep=3.5pt},  blank/.style={draw=none,fill=none,ultra thick,inner sep=0.001pt},scale=4.5]
      \node [inner] (v1)  {\large $1$};
      \node [inner,below=0.6cm of v1] (v2) {\large $2$};
      \node [inner,right=2.2cm of v1] (v3) {\large $3$};
      \node [inner,right=2.2cm of v2] (v4) {\large $4$};
      \draw[very thick] (v1) -- (v2); 
      \draw[very thick] (v1) -- (v3); 
      \draw[very thick] (v2) -- (v4); 
      \draw[very thick] (v2) -- (v3); 
      \draw[very thick] (v4) -- (v3);       
      \node[blank,above=0.0cm of v1] {  ${\color{red}\lbrace \textrm{blue, red} \rbrace}$};
      \node[blank,below=0.0cm of v2] {  ${\color{red}\lbrace \textrm{orange, blue} \rbrace}$};
      \node[blank,above=0.0cm of v3] {  ${\color{red}\lbrace \textrm{red, orange} \rbrace}$};
      \node[blank,below=0.0cm of v4] {  ${\color{red}\lbrace \textrm{blue, red} \rbrace}$};
      \node[blank,below left=0.02cm and 0.0cm of v1] {  $e_1$};
      \node[blank,below left=0.32cm and -0.25cm of v1] {  $\lbrace {\color{blue}1,2}\rbrace$};
      \node[blank,above right=-0.15cm and 0.75cm of v1] {  $\lbrace {\color{blue}4,1} \rbrace$};
      \node[blank,above right=0.25cm and 1.0cm of v1] {  $e_2$};
    \node[blank,below right=-0.15cm and 1.1cm of v2] {  $e_3$};
      \node[blank,below right=0.1cm and 0.85cm of v2] {  $\lbrace {\color{blue}1,3} \rbrace$};
      \node[blank,below right=0.02cm and 2.95cm of v1] {  $e_4$};
      \node[blank,below right=0.32cm and 2.65cm of v1] {  $\lbrace {\color{blue}2,3} \rbrace$};
    \node[blank,below left=0.10cm and 1.35cm of v3] {  $\lbrace {\color{blue}3,2} \rbrace$};
      \node[blank,below left=-0.20cm and 1.65cm of v3] {  $e_5$};
    \end{tikzpicture}

    \caption{{An example of a graph-temporal trajectory of length two, with the red tuples that indicates the node labels, and the blue tuples that indicates edge labels. The first and second items in the labels are for time index 1 and 2, respectively.}}
\label{fig:gtl_example}
\end{figure}

{
\begin{example}
Referring back to the example in Fig.~1, we show an example with $4$ police officers in Fig.~\ref{fig:gtl_example}. The node and edge labels are from a graph-temporal trajectory $g$ of length 2 (with time indices $1$ and $2$). The node labels represent the intersection that the corresponding police officer is monitoring, and the edge labels represent the distance between two neighboring officers. The atomic node proposition $ \pi=(x = \textrm{blue} \vee \textrm{orange})$ is satisfied by $g$ at nodes $1$, $2$, and $4$ at time index $1$, and at nodes $2$ and $3$ at time index $2$. The edge proposition $\rho = (y \leq 2)$ is satisfied by $g$ at $e_1$, $e_3$, and $e_4$ at time index $1$, and at $e_1$, $e_2$, and $e_5$ at time index $2$.
\end{example}}

\begin{definition}[Neighboring operation]
	\label{N}
	Given a graph-temporal trajectory $g=(s,y)$ on a graph $G$, a subset $V'\subseteq V$ of nodes and an edge proposition $\rho$, we define the \textit{neighboring operation} $\bigcirc_{\rho}$ : $2^{V}\times\mathbb{T}\rightarrow 2^{V}\times\mathbb{T}$ as%
	\begin{align*}
	&\bigcirc_{\rho}(V',t) =\big(\{v \;|\; \exists v'\in V', \exists e\in E, s(e)=\{v',v\},\\
	&(g,e,t)\models\rho\},t\big).
	\end{align*}%
	The set $\bigcirc_{\rho}(V',t)$ is the set of nodes that can be reached from $V'$ through an edge where the edge proposition $\rho$ is satisfied by $g$ at time index $t$. We can apply neighboring operations successively.
\end{definition}

A graph-temporal trajectory $g$ at a node $v$ and at a time index $t$ satisfies $\exists^{N}(\bigcirc_{\rho_{n}}\cdots \bigcirc_{\rho_{1}})\varphi$ if there exist at least $N$ nodes in $(\bigcirc_{\rho_{n}}\cdots \bigcirc_{\rho_{1}})(v,t)$ where the formula $\varphi$ is satisfied by $g$ at $t$. 
A graph-temporal trajectory $g$ satisfies a GTL formula $\varphi$ at a node $v$, denoted as $(g, v)\models\varphi$, if $g$ satisfies $\varphi$ at node $v$ at time index 0. 
The syntax of the GTL forumulas is given in~\cite{zhe_GTL}.

\begin{definition} 
	A deterministic finite automaton (DFA) is a tuple $\mathcal{A}=(\mathcal{Q},\qinit,\Sigma,\delta,\Acc)$ where
	with a finite set $\mathcal{Q}=\{q_0, q_1, \dots, q_{K-1}\}$ of states, an initial state $\qinit$, an alphabet $\Sigma$, a transition relation $\delta:\mathcal{Q}\times\Sigma\rightarrow\mathcal{Q}$, and a set of accepting states $\Acc\subseteq2^{\mathcal{Q}}$\cite{KupfermanVardi2001}. 
	\label{sw1}                               
\end{definition}   

For the co-safe (resp. safe) GTL formulas that we consider, we can construct a DFA $\mathcal{A}^{\varphi,v}$ (resp. $\mathcal{A}^{\lnot\varphi,v}$) over $\mathcal{AP}$ that only accepts the graph-temporal trajectories that satisfy (resp. violate) the GTL formula $\varphi$ at node $v$~\cite{zhe_GTL}.   

\vspace{-0.4cm}
\added{\begin{definition}[Time horizon of GTL formulas]
The time horizon $T_{\varphi} \in \mathbb{T}$ of a GTL formula is the
minimum time length of a graph-temporal trajectory $\rho$ to evaluate both
its satisfaction and its violation with respect to $\varphi$.
\end{definition}}%

\vspace{-0.4cm}
\added{\begin{example}
For a GTL formula $\varphi_1=X(\Diamond_{[0,5]}(\exists^2 x\geq 2))$, the time horizon of $\varphi_1$ is $T_{\varphi_{1}}=6$. The time horizon of $\varphi_2=\Box_{[0, 10]}\left(\Diamond_{\leq 3}\left(\exists^2 \bigcirc_{y\leq 4} x\geq 3 \right)\right)$ is $T_{\varphi_{2}}=13$.
\end{example}}%

\vspace{-0.3cm}
\section{Problem Formulation}\label{GTL_sec}

We synthesize a policy for each agent that satisfies the GTL specification. 
We construct a factored MDP that captures all trajectories of a factored MDP $\mdp$ satisfying a GTL formula $\varphi$ by taking the product of $\mdp$ and the DFA $\mathcal{A}^{\varphi}$ corresponding to the GTL formula $\varphi$. 
We represent the GTL specifications as reachability specifications on the product factored MDP. 




\begin{definition}[Product factored MDP]
Let $\MdpInit$ be a factored MDP and $\DraInit$ be a DFA. 
The product factored MDP is given by a tuple $\ProductInit$ with a finite set $\Sp=S \times Q$ of states, an initial state $\sinitp=(\sinit,\qinit)\in \Sp$ that satisfies $q=\delta(\qinit,\mathcal{L}(\sinitp))$, a finite set $\Act$ of actions, a labeling function $\Labelp((s,q))=\{q\}$, a transition function $\probmdpp_i((s,q), \act, (s',q'))= \probmdp_i(s,\act,s')\;  \text{if} \quad q'=\delta(q,\mathcal{L}(s'_i))$, and  $\probmdpp_i((s,q), \act, (s',q'))=0 \; \allowbreak \text{otherwise}$, a reward function $\Rp$ that satisfies $\Rp(s,q,\act)=R(s,\act)$ for $s \in S$ and $\act \in \Act$, the acceptance condition $\Accp=\{\Acc_1^\mathrm{p},\ldots,\Acc_k^\mathrm{p} \}$, where $\Acc_i^\mathrm{p}=\Sp_{i}\times \Acc_i$ for all $\Acc_i\in \Acc$. 
We assume all accepting states are absorbing.\looseness=-1
\end{definition}


%
 
The policy $\schedp$ on a product factored MDP $\mdpp$ satisfy the GTL specification $\varphi$ with a probability of at least $\lambda$ at node $v$ and at a time index $t$, i.e., $\mathbb{P}_{\mdpp_{\schedp}}((\varrho_{\schedp},v,t)\models \varphi)\geq \lambda$, if and only if the trajectories of the product factored MDP $\mdpp$ under the policy $\sched$ reach some accepting states in $\mdpp$ with a probability of at least $\lambda$~\cite{BK08}.

\begin{definition}[Occupancy measure]
The occupancy measure $o^\sched_i$ of a policy $\sched_i$ for a set of neighboring agents $N(i)$ of a factored MDP $\mdp$ over \added{the time horizon $T_\varphi$} is%
\begin{align}
&   \displaystyle o^{\sched_i}_i(\hat{s}_{N(i)},\hat{\act}_{N(i)})=
\mathbb{E}\Big[ \sum\nolimits_{t=0}^{\added{T_\varphi}}\mathbb{P}(\hat{s}_{N(i)}(t)=\hat{s}_{N(i)},\nonumber\\
&\hat{\act}_{N(i)}(t)=\hat{\act}_{N(i)}|\hat{s}_{N(i)}(0)=\hat{s}^{I}_{N(i)})\Big],
\end{align}%
where $\hat{s}_{N(i)}(t)=\lbrace s_{j_1}(t), \ldots ,s_{j_{|N(i)|}}(t) \rbrace \in S_{N(i)}$ and $\hat{\act}_{N(i)}(t) \allowbreak= \allowbreak \lbrace  \act_{j_1}(t), \allowbreak\ldots, \act_{j_{|N(i)|}}(t) \rbrace \allowbreak\in \Act_{N(i)}$ denote the state and action of the agent $v_i$ and the neighboring agents in $\mdp$ at time index $t$. 
The equality $\hat{s}_{N(i)}(t)=\hat{s}_{N(i)}$ means all elements of $\hat{s}_{N(i)}(t)$ and $\hat{s}_{N(i)}$ are same. The value of  $o^{\sched_i}_i(\hat{s}_{N(i)},\hat{\act}_{N(i)})$ is the expected number of taking the action ${\act}_{j_k}$ at the state ${s}_{j_k}$ for all $k \in N(i)$ under the policy $\sched_i$.
\end{definition}%



\subsection{Problem Statement} %
We state the policy synthesis problem of a factored MDP subject to GTL specifications. 
We synthesize a policy that generates the trajectories that satisfy the given GTL specification with at least a desired probability while maxiziming the obtained reward.
To this end, we synthesize a policy $\schedp \in \Sched_{\mdpp}$ that reaches the accepting states in $\mdpp$ with a probability of at least $\lambda$.%
\begin{problem}
Given a factored product MDP $\mdpp$, a planning horizon $T$, compute a policy $\schedp \in \Sched_{\mdpp}$ that solves%
\begin{align}
    & \displaystyle \underset{\sched^\mathrm{p} \in \Sched_{\mdpp}}{\textnormal{maximize}}\quad \mathbb{E}\left[ \sum\nolimits_{t=0}^{\added{T_\varphi}}R(s(t), \act(t))\right]\\
    & \displaystyle \textnormal{subject to} \quad \mathbb{P}_{\mdpp_{\schedp}}( (\varrho_{\schedp},v,k) \models \varphi)\geq \lambda,
\end{align}%
where $\mathbb{P}_{\mdpp_{\schedp}}((\varrho_{\schedp},v,k)\models \varphi)$ denotes the probability of satisfying the GTL specification $\varphi$ with the trajectory $\varrho_{\schedp}$ for some agents $v \in \hat{V}$. $\hat{V}$ is the set of agents that have a GTL specification that is required to be satisfied at the time index $k$ in the factored product MDP $\mdpp$ under the policy $\schedp$. Without loss of generality, we assume $k=0.$
\end{problem}%
\section{Policy Synthesis}%
We now describe the proposed approaches to synthesize a policy for each agent to solve Problem~1. We first give two centralized formulations based on a linear programming problem. 
The first formulation computes a policy for each agent that is a function itself and its neighboring agent's states.
The second formulation computes a policy for each agent that is only a function of their states.
We then develop a distributed approach based on centralized formulations. 
\subsection{Centralized Approach}%
In this section, we propose a linear programming problem (LP) for solving Problem~1. 
Our solution is based on the formulation to compute a policy that maximizes the expected reward while satisfying a temporal logic specification \added{$\varphi$} in an MDP~\cite{puterman2014markov,forejt2011quantitative}.
We denote $\bar{S}$ be the set of all non-accepting states in $\mdp$, i.e., that are not in $Acc$. Then, we define the variables of the dual LP formulation as:%
		\begin{itemize}
			\item \added{$o(s,\act,t)\in [0,1]$ for $s\in \bar{S}, \act \in \Act,$ and $t \in T_\varphi$ gives the expected number of taking action $\act$ in a state $s$ at time index $t$, which also defines the occupancy measure of each state and action.}
			\item \added{$o(s, t) \in [0, 1]$ for $s \in Acc$ and is the probability of reaching an accepting state $s$ at time index $t$.}	\end{itemize}
We note that we define the variables $o(s,t)$ in the interval $[0, 1]$ instead of $[0, \infty)$, as they represent the probability of reaching an accepting state $s \in Acc$. 
We refer to~\cite{forejt2011quantitative,savas2019entropy} for further explanation of the domain of the variables.

The LP is given by%
\begin{align}
		   \displaystyle	&\text{maximize} \quad   \displaystyle \sum_{s\in \bar{S}}\sum_{\act\in\Act}\added{\sum_{t \in \bar{T}_\varphi}\added{o(s,\act,t)}}R(s,
		   act)\label{eq:policylp:obj}\\
			&\text{subject to} \quad \forall s\in \bar{S}, \added{\forall t \in \bar{T}_{\varphi}-1},\label{eq:policylp:welldefined_sched}
\\
		& \displaystyle	\sum\limits_{\act\in\Act}\added{o(s,\act,t+1)} -\mu(s)= \sum\limits_{s'\in \bar{S}}\sum\limits_{\act\in\Act}\probmdp(s|s',\act)\added{o(s',\act,t)}, \nonumber	\\
				&	\forall s\in Acc, \added{\forall t \in \bar{T}_{\varphi}-1},\label{eq:policylp:mec_sched}
				\\			&\displaystyle  \added{o(s,t+1)}-\mu(s)= \sum_{s'\in \bar{S}}\sum_{\act\in\Act}\probmdp(s|s',\act)\added{o(s',\act,t)}+\added{o(s,t)}\nonumber\\
		   \displaystyle		& \displaystyle 	 \sum_{s \in Acc}\added{o(s,\bar{T}_\varphi)} \geq \lambda,\label{eq:policylp:probthresh}
\end{align}%
\added{where $\bar{T}_{\varphi}=\lbrace{0,1,\ldots,\bar{T}_\varphi\rbrace}$, and $\mu(s)=1$ if $s=\sinit$ and $t=0$, and  $\mu(s)=0$ otherwise.} 
The constraints~\eqref{eq:policylp:welldefined_sched} and~\eqref{eq:policylp:mec_sched} are referred to as \emph{flow} constraints~\cite{puterman2014markov}, and ensures that the transitions between each states in the MDP is well-defined.
The constraint~\eqref{eq:policylp:probthresh} ensures that the specification $\varphi$ is satisfied with a probability of at least $\lambda$. 

Given an optimal solution $o$ for the LP in~\eqref{eq:policylp:obj}--\eqref{eq:policylp:probthresh}, an optimal policy can be computed by%
\begin{align}
   \displaystyle \sched(s,\act,t)= \dfrac{o(s,\act,t)}{   \sum_{\act'\in\Act}o(s,\act',t)}.\label{eq:occupmeasure}
\end{align} %
and $o$ is the occupancy measure of $\sched$~\cite{puterman2014markov,forejt2011quantitative}.  
\added{In the remainder of the paper, we drop the dependency of $t$ from the occupancy variables in the formulation of constraints for brevity.}

\subsection{Synthesis of Neighboring Policies}

We now describe the centralized approach for the policy synthesis problem for factored MDPs subject to GTL specifications. 
Let $\bar{S}^{\mathrm{p}}_{N(i)}$ be the set of all states in the product factored MDP $\mdpp$ that are not in $\Acc^\mathrm{p}_i$ for each agent $v_i$. 
We define the variables of the LP for policy synthesis as 
		\begin{itemize}
			\item $o_{i}(\hat{s}_{N(i)},\hat{\act}_{N(i)})\in [0,\infty)$ for each set of neighboring states $\hat{s}_{N(i)} \in \bar{S}^{\mathrm{p}}_{N(i)}$ and actions $\hat{\act}_{N(i)}\in\Act_{N(i)}$ defines the occupancy measure of a state-action pair for $\sched_i$.
			\item $o_{i}(\hat{s}_{N(i)}) \in [0, 1]$ for each state $\hat{s}_{N(i)} \in \Acc^p_i$ defines the probability of reaching an accepting state $s\in \Acc^p_i$.
		\end{itemize}

The objective of the LP is given by%
\begin{align}
\text{maximize} \quad    \displaystyle& \displaystyle \sum\limits_{i=1}^{M}  \sum\limits_{\hat{s}_{N(i)}\in \bar{S}^{\mathrm{p}}_{N(i)}}\sum\limits_{\hat{\act}_{N(i)}\in\Act_{N(i)}}\nonumber\\ & \displaystyle o_i(\hat{s}_{N(i)},\hat{\act}_{N(i)})R_i(\hat{s}_{N(i)},\hat{\act}_{N(i)}).\label{eq:graph_mdp_obj}
\end{align}%
For each agent $v_i \in V$, and state $\hat{s}_{N(i)} \in \bar{S}^{\mathrm{p}}_{N(i)}$, the constraints%
\begin{align}
    & \displaystyle \sum\limits_{\hat{\act}_{N(i)}\in\Act_{N(i)}}o_i(\hat{s}_{N(i)},\hat{\act}_{N(i)}) -\mu(\hat{s}_{N(i)})=\label{eq:graph_mdp_flow}\\
    &\displaystyle \sum\limits_{\hat{s}'_{N(i)}\in \bar{S}^{\mathrm{p}}_{N(i)}}\sum\limits_{\hat{\act}_{N(i)}\in\Act_{N(i)}}\probmdp_i^{\mathrm{p}}(\hat{s}_{N(i)}|\hat{s}'_{N(i)},\hat{\act}_{N(i)})o_i(s'_i,\hat{\act}_{N(i)})\nonumber
\end{align}%
denote the flow constraints, similar to the constraints~\eqref{eq:policylp:welldefined_sched}.
For each agent $v_i \in V$, and state $\hat{s}_{N(i)} \in \Acc^p_i$, the constraints%
\begin{align}
    & \displaystyle o_i(\hat{s}_{N(i)}) -\mu(\hat{s}_{N(i)})=\label{eq:graph_mdp_flow2}\\
    &\displaystyle\sum\limits_{\hat{s}'_{N(i)}\in \bar{S}^{\mathrm{p}}_{N(i)}}\sum\limits_{\hat{\act}_{N(i)}\in\Act_{N(i)}}\probmdp_i^{\mathrm{p}}(\hat{s}_{N(i)}|\hat{s}'_{N(i)},\hat{\act}_{N(i)})o_i(s'_i,\hat{\act}_{N(i)})\nonumber
\end{align}%
\noindent denote the flow constraints for the accepting states, analogous to the constraints~\eqref{eq:policylp:mec_sched}.

For agents $v_i, v_j \in V$ such that $N(i)\cap N(j) \neq \emptyset$, we ensure that the occupancy measure is consistent in the states and actions of agents $v_k \in N(i,j)$. 
Thus, the agents take account of its neighboring agents' occupation measures during the policy computation. 
For each set of states $\hat{s}_{N(i,j)} \in S_{N(i,j)}$ and actions $\hat{\act}_{N(i,j)} \in \Act_{N(i,j)}$, the constraints%
\begin{align}
     & \displaystyle \sum\limits_{\hat{s}_{N(i)}\supseteq \hat{s}_{N(i,j)}}\sum\limits_{\hat{\act}_{N(i)}\supseteq \hat{\act}_{N(i,j)}}o_i(\hat{s}_{N(i)},\hat{\act}_{N(i)}) =\nonumber\\
     &  \displaystyle \sum\limits_{\hat{s}_{N(j)}\supseteq \hat{s}_{N(i,j)}}\sum\limits_{\hat{\act}_{N(j)}\supseteq \hat{\act}_{N(i,j)}}o_j(\hat{s}_{N(j)},\hat{\act}_{N(j)}) \label{eq:graph_mdp_neighbor}
\end{align}%
ensure that the time spent in the set of states $\hat{s}_{N(i,j)}$ and taking the set of actions $\hat{\act}_{N(i,j)}$ is equal for the policies of agents $v_i$ and $v_j$.
Finally, the constraints%
\begin{align}
    \quad & \displaystyle  \sum\nolimits_{\hat{s}_{N(i)} \in A^p_i} o_i(\hat{s}_{N(i)}) \geq \lambda  \label{eq:graph_mdp_spec}  
    \end{align} %
 encode the specification constraints for each agent $v_i \in V$, similar to the constraints~\eqref{eq:policylp:probthresh}. 
 
 We illustrate the constraints~\eqref{eq:graph_mdp_neighbor} by an example.
    \begin{example}
    Consider the factored MDP in Fig.~\ref{fig:graph_mdp} and the agents $v_1$ and $v_2$. $N(1,2)=\lbrace1,2,3\rbrace, N(1\setminus 2)=\emptyset$ and $N(2 \setminus 1)=\lbrace 4 \rbrace$. We add the constraints to ensure that the occupancy measure is consistent for agents $v_1$ and $v_2$
    \begin{align}
        & o_1(\hat{s}_{N(1,2)},\hat{\act}_{N(1,2)})=\nonumber\\
       & \displaystyle  \sum\limits_{s_4 \in S_{N(2\setminus 1)}}\sum\limits_{\act_4 \in \Act_{N(2\setminus 1)}}o_2(\hat{s}_{N(1,2)},s_4,\hat{\act}_{N(1,2)},\act_4).
    \end{align}
for $\hat{s}_{N(1,2)} \in S_{N(1,2)}$ and $\hat{\act}_{N(1,2)} \in \Act_{N(1,2)}$.
\end{example}

 \begin{remark}
\added{The constraints in~\eqref{eq:graph_mdp_neighbor} ensure that for each pair of agents $v_i$ and $v_j$, the resulting occupancy measures of their policies are consistent on states that are in both of the respective sets of states of the neighboring agents of $v_i$ and $v_j$.
Specifically, these constraints ensure that the policy of the agents induces a stationary behavior with respect to neighboring agents.}
 \end{remark}
 
\begin{theorem}
     Let $\mathcal{M}_{\textrm{p}}$ be a product factored MDP, $A_i$ $i=1,\ldots,p$ be the accepting states in $\mathcal{M}_{\textrm{p}}$. For the input ($\mathcal{M}_{\textrm{p}}, A_i )$, an optimal policy $\sched^{\textrm{p},\star}_i$ for each agent $i$ that satisfies the GTL specifications and maximizes the expected reward can be computed by an optimal solution of the LP in~\eqref{eq:graph_mdp_obj}--\eqref{eq:graph_mdp_spec}.
\end{theorem}
 \begin{proof}
 See Appendix B.
 \end{proof}


\subsection{Synthesis of Local Policies}

The number of variables in the previous approach scales exponentially with the number states and agents of the neighboring agents, which may be time-consuming even with a modest number of neighboring agents. 
In this section, we propose restricting the policy of the agents to their local states, and we only let the agents interact through the specification constraints, which also defines the coupling in the LP formulation.
In order to derive this approach, we put restrictions on the transition functions, set of actions, and the reward functions. 
We make the following modifications:

\begin{itemize}
    \item For each agent $i$, we define the transition function of agent $i$ as $\mathcal{P}_i \colon S_i \times \Act_i \rightarrow \Distr(S_i)$. 
    \item For each agent $i$, we define the reward function of agent $i$ as $R_i \colon S_i \times \Act_i \rightarrow \mathbb{R}_{\geq 0}.$
    \item For each agent $i$, we define the policy of agent $i$ as $\sched_i \colon S_i \rightarrow \Distr(\Act_i)$. 
\end{itemize}

Note that all of these modifications are made to restrict the agent $i$'s local behavior as a function of the agent $i$. 
However, we emphasize that we do not make any modifications to how we describe the tasks, meaning, we can still achieve multi-agent spatial-temporal tasks described by GTL formulas. 

\added{\begin{remark}
The policy synthesis takes into account the interactions from the neighbors to satisfy the GTL specifications while computing a policy that is only a function of the local states.
Therefore, the resulting behavior of the agents is not decentralized with respect to their own states, as the GTL specifications are functions of the state of the agent and its neighboring agents.
\end{remark}}

We now describe the centralized approach for the policy synthesis problem for factored MDPs subject to GTL specifications with the restricted policy. 
Let $\bar{S}^{\mathrm{p}}_{N(i)}$ be the set of all states in the product factored MDP $\mdpp$ that are not in $\Acc^\mathrm{p}_i$ for each agent $v_i$ and $\bar{S}_i$ be the set of all states in $\Acc_i$ for each agent $v_i$.
Then, we define the variables of the LP for policy synthesis with local policies as follows:
		\begin{itemize}
			\item $o_{i}(s_i,\act_i)\in [0,\infty)$ for each state $s \in \bar{S}_i$ and action $\act_i\in\Act_i$ defines the expected number of taking an action $\act_i$ in the state $s_i$.
			\item $o_{i}(s_i) \in [0, 1]$ for each state $s_i \in \Acc_i$ defines the probability of reaching an accepting state $s_i\in \Acc_i$.
						\item $o_{i}(\hat{s}_{N(i)})\in [0,\infty)$ for each set of neighboring states $\hat{s}_{N(i)} \in \bar{S}^{\mathrm{p}}_{N(i)}$ defines the occupancy measure of a state for $\sched_i$.
			\item $o_{i}(\hat{s}_{N(i)}) \in [0, 1]$ for each state $\hat{s}_{N(i)} \in A^p_i$ defines the probability of reaching an accepting state $s\in A^p_i$.
		\end{itemize}
		
The objective of the LP is%
\begin{align}
\text{maximize} \quad    \displaystyle& \displaystyle \sum\limits_{i=1}^{M}  \sum\limits_{s_i \in \bar{S}_i}\sum\limits_{\act_i\in\Act_i} \displaystyle o_i(s_i, \act_i)R_i(s_i,\act_i).\label{eq:graph_mdp_local_obj}
\end{align}%
For each agent $v_i \in V$, and state $s_i \in \bar{S}_i$, the constraints %
\begin{align}
    & \displaystyle \hspace{-0.05cm} \sum\limits_{\act_i \in \Act_i} \hspace{-0.10cm} o_i(s_i,\act_i) -\mu(s_i)=\hspace{-0.05cm}\label{eq:graph_mdp_local_flow1}\sum\limits_{s'_i\in \bar{S}_i}\sum\limits_{\act_i \in \Act_i}\hspace{-0.12cm}\probmdp_i(s_i|s'_i,\act_i)o_i(s'_i,\act_i)\raisetag{0.8\baselineskip}
\end{align}%
\noindent denote the flow constraints for the non-accepting states, analogous to the constraints~\eqref{eq:policylp:welldefined_sched}.

For each agent $v_i \in V$, and state $s_i \in \Acc_i$, the constraints%
\begin{align}
    & \displaystyle o_i(s_i) -\mu(s_i)=\label{eq:graph_mdp_local_flow2}\sum\limits_{s'_i\in \bar{S}_i}\sum\limits_{\act_i \in \Act_i}\probmdp_i(s_i|s'_i,\act_i)o_i(s'_i,\act_i)
\end{align}%
\noindent denote the flow constraints for the accepting states, analogous to the constraints~\eqref{eq:policylp:mec_sched}.

\added{For each agent $v_i \in V$ and state $s_i \in \bar{S}_i$, the  constraints%
\begin{align}
    & \displaystyle \sum\nolimits_{\act_i \in \Act_i} o_i(s_i,\act_i)=\label{eq:graph_mdp_local_neighbor}\sum\nolimits_{\hat{s}_{N(i)}\in s_i \cup \bar{S}^{\mathrm{p}}_{\hat{N}(i)}} o_i(s_i,\hat{s}_{\hat{N}(i)})
\end{align}%
\noindent where $\bar{S}^{\mathrm{p}}_{\hat{N}(i)}$ denotes the of neighboring states in $\bar{S}^{\mathrm{p}}_{\hat{N}(i)}$ without including the states $s_i \in S_i$. 
The constraints ensure that the occupation measure in the neighboring states for agent $v_i$ is consistent with the local policy.}

For agents $v_i, v_j \in V$ such that $N(i)\cap N(j) \neq \emptyset$, we ensure that the occupancy measure is consistent in the states of agents $v_k \in N(i,j)$, similar to the previous formulation. 
Thus, the agents take account of its neighboring agents' occupation measures during the policy computation. 
\added{For each state $\hat{s}_{N(i,j)}\in S_{N(i,j)},$ the  constraints%
\begin{align}
     &\displaystyle \sum_{\hat{s}_{N(i/j)}\in S_{N(i,j)}}o_i(\hat{s}_{N(i,j)},\hat{s}_{N(i/j)})=\nonumber\\
    &\displaystyle \sum_{\hat{s}_{N(j/i)}\in S_{N(i,j)}}o_j(\hat{s}_{N(i,j)},\hat{s}_{N(j/i)}) \label{eq:graph_mdp_local_neighbor_occup}
\end{align}%
ensure that the time spent in each state $\hat{s}_{N(i,j)}\in S_{N(i,j)}$ is equal for the policies of agents $v_i$ and $v_j$.
\begin{remark}
The constraints in~\eqref{eq:graph_mdp_local_neighbor_occup} ensure that the behavior of each agent is stationary, similar to~\eqref{eq:graph_mdp_neighbor}.
Additionally, the value of $o_i(\hat{s}_{N(i,j)},\hat{s}_{N(i/j)})$ can also be interpreted as the joint probability of each agent in $N(i,j)$ occupying their respective local state in $\hat{s}_{N(i,j)}$.
\end{remark}
}

Finally, the constraints%
\begin{align}
    \quad & \displaystyle  \sum\nolimits_{\hat{s}_{N(i)} \in A^p_i} o_i(\hat{s}_{N(i)}) \geq \lambda  \label{eq:graph_mdp_local_spec}  
    \end{align}%
 encode the specification constraints for each agent $v_i \in V$, similar to the constraints~\eqref{eq:policylp:probthresh}.%
 
\added{\begin{lemma}[Feasibility of the consistency constraints]
 Given local policies for each agent $v_i \in V$, a feasible solution to the constraints in~\eqref{eq:graph_mdp_local_neighbor_occup} can always be computed. \end{lemma}
 \begin{proof}
 See Appendix B.
 \end{proof}
 The proof of Lemma 1 relies on the fact that given local policies for each agent $v_i \in V$, the behavior of each agent is independent of other agents. 
 Therefore, we can compute the time spent in the neighboring states by computing the occupancy measure of their local states for each agent.}

 \subsection{Discussion of the Complexity of the Proposed Methods}
We now discuss the differences between the formulations in~\eqref{eq:graph_mdp_obj}--\eqref{eq:graph_mdp_spec} and in~\eqref{eq:graph_mdp_local_obj}--\eqref{eq:graph_mdp_local_spec}.
In the first formulation, the number of variables for agent $i$ scales in $\mathcal{O}(|S_{N(i)}||\Act_{N(i)}|)$, and in the second formulation, it scales with $\mathcal{O}(|S_{N(i)}|)$.
The number of terms in the objective for agent $i$ in~\eqref{eq:graph_mdp_obj}  is $\mathcal{O}(|S_{N(i)}||\Act_{N(i)}|)$, and $\mathcal{O}(|S_i||\Act_i|)$ in~\eqref{eq:graph_mdp_local_obj}.
Similarly, the number of flow constraints in~\eqref{eq:graph_mdp_flow}--\eqref{eq:graph_mdp_neighbor} is $\mathcal{O}(|S_{N(i)}||\Act_{N(i)}|)$, and $\mathcal{O}(|S_i||\Act_i|)$ in~\eqref{eq:graph_mdp_local_flow2}--\eqref{eq:graph_mdp_local_neighbor_occup}.

\begin{remark}
\added{Computing policies for each agent $v_i$ as a function of its own states instead of neighboring states significantly improves the running time of the algorithm, both theoretically and empirically.
However, this restriction may result in a policy that incurs a lower reward.}
\end{remark}

\looseness=-1

The LPs in~\eqref{eq:graph_mdp_obj}--\eqref{eq:graph_mdp_spec} and~\eqref{eq:graph_mdp_local_obj}--\eqref{eq:graph_mdp_local_spec} compute a policy for each agent $v_i$ that satisfies the GTL specification and maximizes the expected reward. 
However, the scalability of solving the LP with a centralized algorithm can be challenging if the number of agents $M$ is large. 
In the next section, we propose a distributed approach that runs in time that is linear in $M$.

    \subsection{Distributed Approach}

In this section, we discuss how we can solve the LP in~\eqref{eq:graph_mdp_local_obj}--\eqref{eq:graph_mdp_local_spec} in a distributed manner. 
We utilize alternating direction method of multipliers (ADMM)~\cite{gabay1975dual,boyd2011distributed} to solve a large-scale factored MDP synthesis problem by decomposing them into a set of smaller problems. 
The iterations for ADMM does not necessarily converge to an optimal solution with $M>2$ agents~\cite{chen2016direct}. 
Therefore, we pose the multi-block problem into an equivalent two-block problem, and apply the \emph{primal-splitting ADMM} algorithm~\cite{wang2013solving} to the equivalent problem.



\subsubsection{Primal-Splitting ADMM}

The LP in~\eqref{eq:graph_mdp_local_obj}--\eqref{eq:graph_mdp_local_spec} with $M$ agents can be written as following optimization problem%
\begin{align}
\text{minimize} & \displaystyle\quad ~\sum\nolimits_{i=1}^M f_i(o_i)\label{eq:multi_block_obj} \\
\text{subject to} &\displaystyle\quad ~\sum\nolimits_{i=1}^M B_io_i =0,\label{eq:multi_block_cons}
\end{align}%
where $f_i(o_i)$ is the negative of the objective in~\eqref{eq:graph_mdp_local_obj} and encodes the constraints in~\eqref{eq:graph_mdp_local_flow1}--\eqref{eq:graph_mdp_local_neighbor} and \eqref{eq:graph_mdp_local_spec} in their indicator form for each agent $v_i$. 
The constraints in~\eqref{eq:multi_block_cons} depict the constraints in~\eqref{eq:graph_mdp_local_neighbor_occup} and the matrices $B_i$ encodes the coefficients in~\eqref{eq:graph_mdp_local_neighbor_occup}. 
We note that a similar construction can be done for the synthesis problem with neighboring policies.

\begin{algorithm}[t]
\hspace{-0.30cm}Initialize: $\displaystyle o^0_i$ and $\displaystyle\nu^0_i~(i=1,2,\ldots, M)$\;
\hspace{-0.30cm}{\For{$k=0,1,\ldots,I$}{
\hspace{-0.2cm}\For{$\displaystyle i=1,2,\ldots,M$}{
\hspace{-0.2cm}$\displaystyle z_i^{k+1} \gets - \frac{1}{M}\left(\sum\nolimits_{i=1}^M B_i o_i^k  - \dfrac{\nu_i^k}{\beta} \right) +  B_i o_i^k  - \dfrac{\nu_i^k}{\beta} .$
  
\hspace{-0.2cm}$\displaystyle o_i^{k+1} \hspace{-0.015cm}\gets\hspace{-0.004cm}\underset{o_i}{\text{argmin}} f_i(o_i)+\hspace{-0.000cm}\displaystyle\frac{\beta}{2}\hspace{-0.000cm}\left\|B_io_i-z_i^{k+1}-\dfrac{\nu^k_i}{\beta}\right\|_2^2.$
\hspace{-0.3cm}$\displaystyle\nu_i^{k+1}\gets\nu^k_i -\beta(B_i o_i^{k+1}-z^{k+1}_i).$
}\vspace{0.03cm}
{
\hspace{-0.2cm}$\mathrm{res}_p\gets\sum\nolimits_{i=1}^{M}\Vert B_io^k_i-z^k_i\Vert_2^2.$\\\vspace{0.09cm}
\hspace{-0.2cm}$\mathrm{res}_d\gets\sum\nolimits_{i=1}^{M}\beta B_i\Vert o^k_i-o^{k-1}_i\Vert_2^2.$
}\\
\hspace{-0.2cm}\If{$\mathrm{res}_p\leq \gamma$ \textbf{ and } $\mathrm{res}_d\leq \gamma$}
{\textbf{return} $o_i$ for $i=1,2, \ldots, M.$}}
\textbf{return} $o_i$ for $i=1,2, \ldots, M.$}
\caption{Distributed Synthesis for GTL Specifications with Primal-Splitting ADMM} 
\label{Parallel_N}
\end{algorithm}

We introduce a set of auxiliary variables $z_i, i=1,\ldots,M$ and write the optimization problem in~\eqref{eq:multi_block_obj}--\eqref{eq:multi_block_cons} as%
\begin{align}
\text{minimize} & \quad \displaystyle~\sum\nolimits_{i=1}^M f_i(o_i)\label{eq:multi_block_obj1} \\
\text{subject to} &\quad \displaystyle~B_io_i =z_i,\quad i=1,\ldots,M,\label{eq:multi_block_cons1}\\
&\quad \displaystyle \sum\nolimits_{i=1}^M z_i=0.\label{eq:multi_block_cons2}
\end{align}%
The optimization problems in~\eqref{eq:multi_block_obj}--\eqref{eq:multi_block_cons} and in~\eqref{eq:multi_block_obj1}--\eqref{eq:multi_block_cons2} are equivalent in the sense that they share the same optimal solution set for the variables $o_i, i=1,\ldots,M$.

Algorithm~\ref{Parallel_N} solves the optimization problem using primal-splitting ADMM~\cite{wang2013solving}, where $\beta>0$ is the algorithm parameter, and $\nu_i$ is the dual variable for the constraints in~\eqref{eq:multi_block_cons1}. The proposed method achieves an $O(1/k)$ convergence rate after $k$ iterations~\cite{wang2013solving}. 

{The primal residual for the primal feasibility is given by $\mathrm{res}_p$ and the dual residual for the dual feasibility is given by $\mathrm{res}_d$ in Algorithm 1. }
{The primal residual can be seen as the feasibility residual of the policy  to the specification. The dual residual is the optimality residual of the policy to the objective~\cite{boyd2011distributed}.
We stop the algorithm until it runs for $I$ iterations, or if the residuals are below a threshold $\gamma$.}


\subsubsection{Complexity Analysis}
 
The most computationally challenging step of Algorithm 1 is to solve an LP for the $o_i$ update for $i=1,\ldots, M$. 
The number of variables and constraints in each LP for the $o_i$ update is exponential in $N(i)$, and each LP for the $o_i$ can be solved in time polynomial in the number of variables and constraints via interior-point methods~\cite{nesterov1994interior}.
Therefore, the computation time for each $o_i$ update is exponential in $N(i)$, and the computation time of Algorithm 1 is linear in $M$. 

On the other hand, the number of constraints and variables in  problem~\eqref{eq:multi_block_obj}--\eqref{eq:multi_block_cons} is linear in $M$ and exponential in $N(i)$. 
Therefore, if we solve the optimization problem in~\eqref{eq:multi_block_obj}--\eqref{eq:multi_block_cons} by an interior point method algorithm, then the overall complexity of the algorithm will be \emph{polynomial}, or typically cubic in $M$, and the computation will be challenging for a large number of agents. 
In the numerical examples, we demonstrate that the running time for solving the optimization problem in~\eqref{eq:multi_block_obj}--\eqref{eq:multi_block_cons} directly does not scale linearly in $M$.%
\section{Numerical Examples}
We demonstrate the proposed approaches on three domains: (1) disease management in crop fields, (2) search and rescue, (3) urban security. The simulations were performed on a computer with and Intel Core i7-8665u 1.90 GHz processor and 16 GB of RAM with Gurobi 9.0~\cite{gurobi} as the LP solver.%
\subsection{Disease Management in Crop Fields}
We consider the policy synthesis of a factored MDP for disease management in crop fields, discussed in~\cite{sabbadin2012framework}.
If a crop field is contaminated, the yield of that field decreases, and it can infect its neighbors.
If a field is left fallow, it has a probability of $\xi$ recovering from contamination. 
The decisions for each field at each year for the field $v_i$ are ($\Act_i = \lbrace 1, 2\rbrace$): cultivate normally ($\act_i = 1$) or leave fallow ($\act_i = 2$). 

The problem is then to choose the optimal policy to maximize the expected yield while ensuring that the fields are not contaminated with a high probability. 
We represent the topology of the fields by an undirected graph, where each node in the graph represents one crop field.
We draw an edge between two fields if they share a border and can infect each other.    
We consider a graph where the number of neighbors for each field is four.
Each crop field can be in one of three states: $s_i = 1$ denotes the field is uninfected. $s_i = 2$ and $s_i = 3$ denote different degrees of infection, with $s_i=3$ corresponding to a higher degree of infection. 

The probability that a field transitions from state $s_i= $ to state $s_i =2$ or $s_i=2$ to state $s_i=3$ with $\act_i = 1$ is $\probmdp_i = \probmdp_i (\epsilon, p, n_i) = \epsilon + (1 - \epsilon) (1 - (1 - p)^{n_i} )$, where $\epsilon$ and $p$ are fixed parameters and $n_i$ is the number of the neighbors of $v_i$ that are infected on a given year. 
If the field $i$ is in state $s_i=3$, then it remains in $s_i=3$ with a probability of $1$ if $\act_i=1$ is taken.
The reward function depends on each field’s state and action. 
An uninfected cultivated field achieves a maximal yield of $r = 10$. 
Otherwise, the yield decreases linearly with the infection level, from the maximal reward $r$ to minimal reward $1 + r/10$ at the state $s_i=3$. 
A field left fallow produces a reward of 1. 
As the transition probabilities between states of a field depend on the neighboring field’s states, the method that computes local policies is not applicable here. 
Therefore, we report the results of the approach that computes neighboring policies.\looseness=-1

\begin{table}[t]
\caption{The average yield with 100 fields with 50\% of the fields having a GTL specification.}
\label{table1}
\centering 
\begin{tabular}{|l|l|l|l|l|}
\hline
 \diagbox{$(p,\xi)$}{$\lambda$}      & $0.9$ & $0.8$ & $0.7$ & $0.6$                  \\ \hline
$(0.1, 0.1)$        & $3.61$         & $ 4.15 $           & $ 4.57$              &       $5.05$                \\ \hline
$(0.2, 0.2)$        & $4.92$        & $ 5.81$          & $6.71$         & $7.13$           \\ \hline
$(0.5, 0.5)$        & $7.05$         & $8.12$       & $8.38$           & $8.43$           \\ \hline
$(0.8, 0.8)$        & $8.60$        & $8.95$          & $8.99$              & $9.00$              \\ \hline
\end{tabular}
\end{table}

Reference~\cite{forster2007optimizing} considers computing an optimal disease control strategy of a crop field.
It determines that the optimal control strategy requires treating most of the contaminated fields by immediately leaving them fallow to eradicate the disease. 
We consider the specification  $\varphi=\Box_{[0, 30]}\left(\neg \Diamond \Box_{\leq 2} d \wedge \neg \Diamond \Box_{\leq 3} \exists^{2} \bigcirc d\right)$, where $d$ means the field is infected, which ensures that a contaminated field is treated immediately.
A field satisfies the specification $\varphi$ if the field is uninfected in two time steps in a row, and there should not be two neighboring fields that are infected three time steps in a row for the next $30$ time steps. 
We require that 50\% of the fields satisfy the specification.
Table~\ref{table1} shows the average yield for different values of $\lambda, p$, and $\xi$. 
The average yield of the fields increases with lower values of $\lambda$ satisfying the specification. 
However, the fields are infected with an increasing likelihood.
Specifically, for $\lambda=0.9$, the average probability of having an infection of the fields with a GTL specification is $0.03$, and the fields without a GTL specification is $0.07$.
\looseness=-1

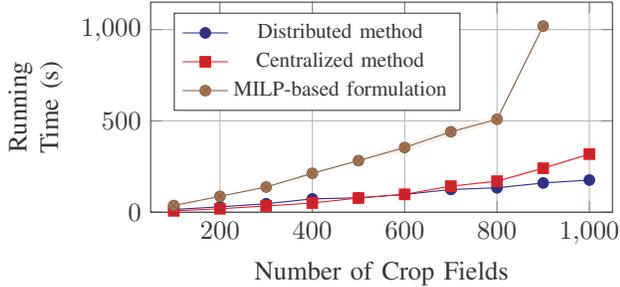
\begin{figure}[t]
\centering
\begin{tikzpicture}
\begin{axis}[%
width=2.417in,
height=1.1in, 
at={(2.167in,0.898in)},
scale only axis,
xmin=50,
xmax=1050,
yminorticks=true,
ymajorticks=true,
xmajorgrids=true,
ymajorgrids=true,
legend style={at={(0.05,0.95)},anchor=north west,font=\fontsize{8}{8.5}\selectfont},
xlabel style={font=\color{white!0!black}},
xlabel={Number of Crop Fields},
ymin=0,
ymax=1150,
yminorticks=true,
ylabel style={font=\color{white!0!black},align=center},
ylabel={Running\\Time (s)},
axis background/.style={fill=white},
title style={font=\bfseries},
title={},
]
\addplot table[x=x,y=y] {admm_output_compile}; 
\addlegendentry{Distributed method}
\addplot table[x=x,y=y] {cent_output_compile}; 
\addlegendentry{Centralized method}
\addplot table[x=x,y=y] {milp_output_compile2}; 
\addlegendentry{MILP-based formulation}

\addplot [name path=upper1,draw=none] table[x=x,y expr=\thisrow{y}+\thisrow{ey}] {admm_output_compile};
\addplot [name path=lower1,draw=none] table[x=x,y expr=\thisrow{y}-\thisrow{ey}] {admm_output_compile};
\addplot [fill=blue!10] fill between[of=upper1 and lower1];
\addplot [name path=upper,draw=none] table[x=x,y expr=\thisrow{y}+\thisrow{ey}] {cent_output_compile};
\addplot [name path=lower,draw=none] table[x=x,y expr=\thisrow{y}-\thisrow{ey}] {cent_output_compile};
\addplot [fill=red!10] fill between[of=upper and lower];
\addplot [name path=upper,draw=none] table[x=x,y expr=\thisrow{y}+5*\thisrow{ey}] {milp_output_compile2};
\addplot [name path=lower,draw=none] table[x=x,y expr=\thisrow{y}-5*\thisrow{ey}] {milp_output_compile2};
\addplot [fill=brown!10] fill between[of=upper and lower];
\end{axis}
\end{tikzpicture}
\caption{\added{The running time of the proposed distributed synthesis method (in Algorithm 1), centralized method, and the MILP-based formulation with a different number of crop fields.}
The shaded region shows the standard deviation of the running times in 20 runs. 
\added{The running time of the decentralized algorithm scales linearly with the number of crop fields. The running time of the MILP-based formulation is larger compared to other methods for a different number of crop fields.}\looseness=-1}
\label{fig:num_agent}
\end{figure}

We also show the scalability of the approach with a different number of crop fields. 
We show the running time of Algorithm~1 and the centralized method for different number of crop fields in Fig.~\ref{fig:num_agent}. 
\added{We also make a comparison of an equivalent mixed-integer linear programming (MILP) formulation which is often used for policy synthesis with metric temporal logic specifications.~\cite{sayan2016}.}
The results in Fig.~\ref{fig:num_agent} show that the running time of the distributed approach is linear with the number of crop fields, unlike the centralized method. 
We observe that Algorithm~1 is slower than the centralized method with fewer crop fields due to solving many smaller LPs iteratively.
\added{Additionally, both the Algorithm~1 and the centralized method significantly outperforms the MILP formulation with increasing number of crop fields.}\looseness=-1
\subsection{Search and Rescue}%

\begin{figure}
    \centering
\includegraphics[width=1.00\linewidth]{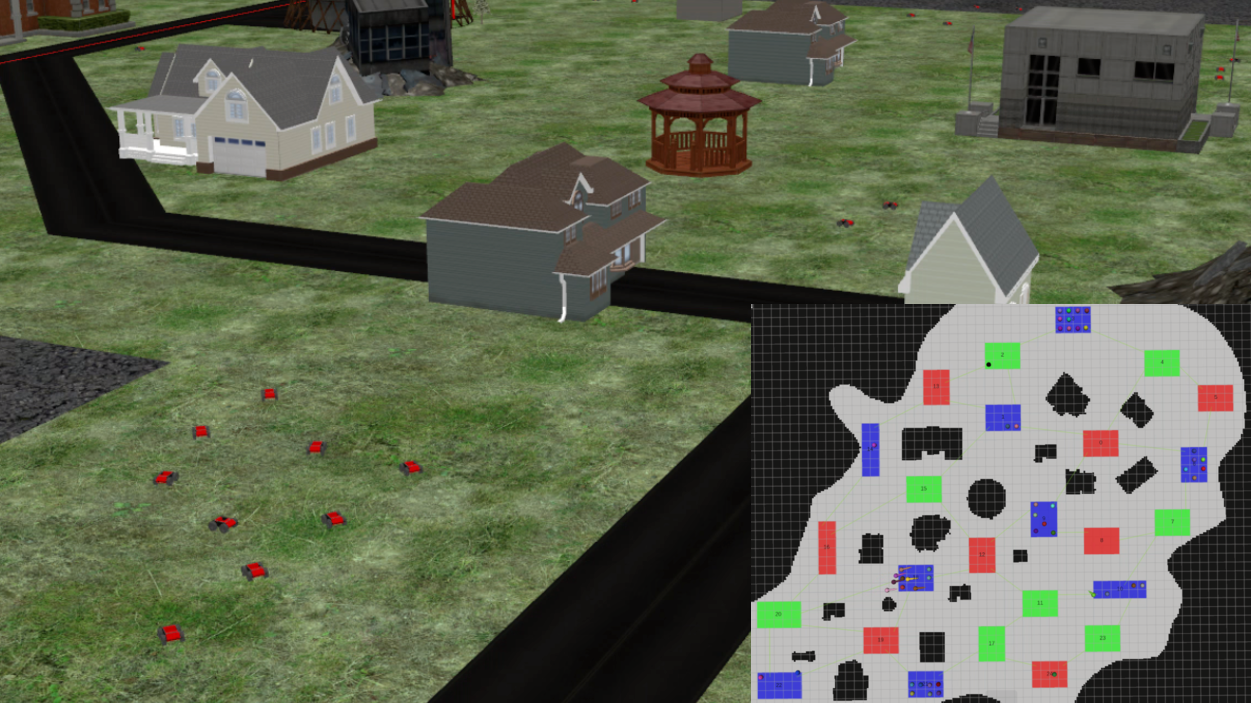}
\caption{Snapshot of the Gazebo environment for the search and rescue mission with $50$ ground robots. We show the locations of the 25 areas in the environment at the bottom right.}
    \label{fig:multiagent}
\end{figure}

In this example, we consider a scenario where a building has collapsed, and 50 robots are tasked to perform search and rescue activities to find victims in a simulated environment in Gazebo, a 3D robotics simulator.
We divide the environment into 25 areas and task the agents to perform search and rescue operations in each area, see Fig.~\ref{fig:multiagent}.
We label the most critical areas as red.
The goal of agents is to accumulate maximum rewards by visiting the red states while satisfying the following specification with a probability of at least $\lambda$:%
\begin{align*}
     \varphi=\Box_{[0,20]} \Big(\textrm{Red} \rightarrow& \big(\Diamond_{\leq 2} (\exists^3 \!\bigcirc_{y\leq 4}\!\textrm{Green}\!\wedge\exists^3\!\bigcirc_{y\leq 4}\!\textrm{Blue})\big)\Big).
\end{align*}
The specification $\varphi$ reads as \textquotedblleft (for any agent) whenever the agent is in a red-colored area, then, for the next two time steps, there are always at least three neighboring agents within distance four in a green-colored area and at least three neighboring agents within distance four in a blue-colored area in the next $20$ time steps\textquotedblright.
The node labels in this problem are the colors of an area, and the edge labels are the distances between each agent.

We consider an example with 50 agents, with each agent having three neighbors in the underlying graph. 
\added{We could only apply the method with local policies on this example, as the method that computes neighboring policies runs out of memory while solving the underlying optimization problem.}

Fig.~\ref{fig:multiagent} shows a snapshot of the Gazebo environment with $50$ ground robots and the locations of the 25 areas in the environment.
\added{We achieve the collision avoidance using optimal reciprocal collision avoidance in the Gazebo simulation~\cite{van2011reciprocal}.}
The computation time of the procedure was 684.14 seconds with $\lambda=0.95$.
We also demonstrate the behavior of the robots in a video\footnote{A video in Gazebo environment can be found in https://bit.ly/3mBFh4W.}.
Initially, all robots are in red-colored areas. 
We observe that the robots maximize the time spent in red areas to incur maximal reward while visiting green- and blue-colored areas to satisfy the GTL specification $\varphi$.
For example, in time indices 2 and 10, most of the robots are in the blue and purple-colored areas to satisfy the GTL specification, even though the robots do not incur reward.\looseness=-1%

\subsection{Urban Security}

We consider an urban security problem, where a criminal is planning his crime, given the number of police officers on the nearby locations~\cite{wu2019reward}. 
Each location should be monitored by a police officer to prevent crimes.
There are $M$ police officers that are tasked to monitor the locations. 
Each police officer coordinates with a sub-group of police officers in monitoring the locations to prevent crimes.
\added{We provide additional details on the results in  Appendix A. 
We make an empirical comparison between computing policies as a function of their own states and the neighboring agent's states.
We show that we can compute policies, on average, two orders of magnitude faster by computing local policies, and the difference between the expected rewards is not significant.}

\section{Conclusions and Future Work}%
We proposed a method for the distributed synthesis of policies for multi-agent systems to satisfy spatial-temporal tasks. 
We express the spatial-temporal tasks in a specification language called graph temporal logic.
For such systems, we decomposed the synthesis problem into a set of smaller problems, one for each agent. 
In the numerical examples, we showed that the algorithm scales to hundreds of agents with hundreds of states per agent.

For future work, we will consider scenarios where the edges of the graph are functions of the agent's local states, and the agents may share the state information with its neighboring agents in certain parts of the state spaces. 

%





\ifCLASSOPTIONcaptionsoff
  \newpage
\fi



%
\bibliographystyle{IEEEtran}  
\bibliography{sample-bibliography}  

%

\section{Appendix~A}

In this appendix, we demonstrate the details for the example on the urban security domain.

\subsection{Urban Security}%

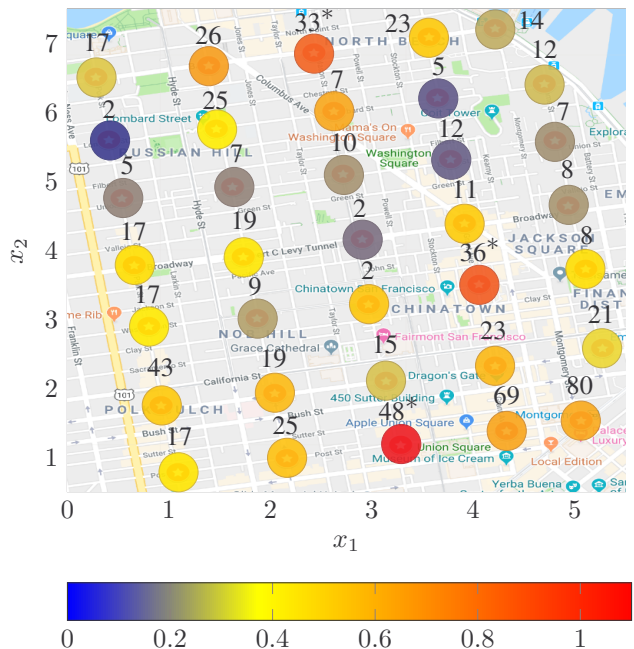
\begin{figure}[t] 
    \centering
\begin{tikzpicture}[remember picture]
\centering
\begin{axis}[
xmin=0, xmax=5.5,
ymin=0.5, ymax=7.5,
xlabel={$x_1$},ylabel={$x_2$},
width=9.0cm,height=8.0cm,colorbar horizontal,point meta min=0.0,
    point meta max=1.1,colorbar style={
        width=7.5cm,
        xtick={0,0.2,0.4,0.6,0.8,1,1.2}}
]
\addplot graphics [includegraphics cmd=\pgfimage,xmin=0, xmax=5.5, ymin=0.5, ymax=7.5] {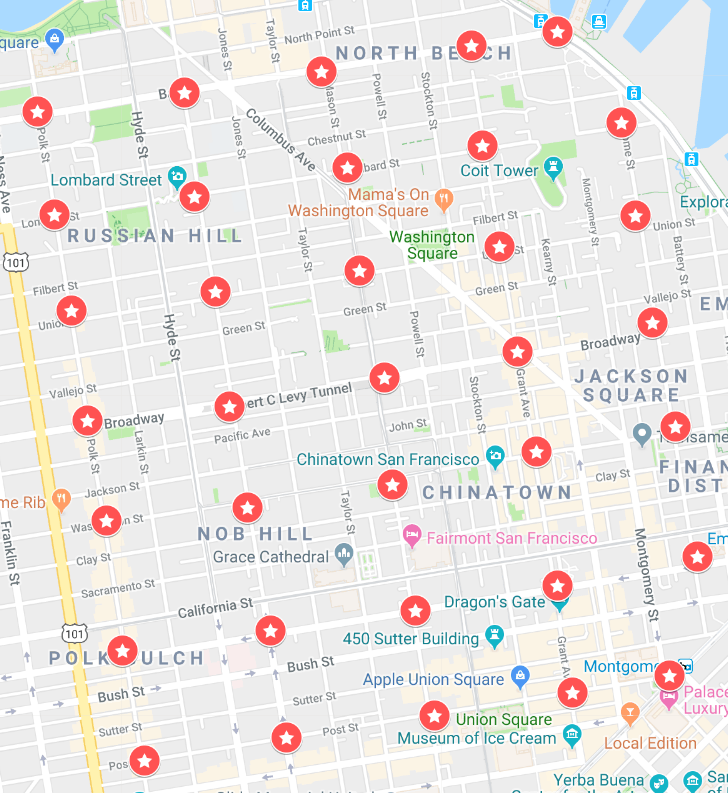};
\node [] at (110,80) {$\text{17}$};
\node [draw=none] at (215,97) {$\text{25}$};
\node [draw=none] at (327,121) {$\text{48}^{\displaystyle \bf{*}}$};
\node [draw=none] at (433,140) {$\text{69}$};
\node [draw=none] at (505,155) {$\text{80}$};
\node [draw=none] at (92,180) {$\text{43}$};
\node [draw=none] at (203,194) {$\text{19}$};
\node [draw=none] at (313,214) {$\text{15}$};
\node [draw=none] at (420,234) {$\text{23}$};
\node [draw=none] at (526,258) {$\text{21}$};
\node [draw=none] at (79,286) {$\text{17}$};
\node [draw=none] at (185,298) {$\text{9}$};
\node [draw=none] at (296,321) {$\text{2}$};
\node [draw=none] at (407,351) {$\text{36}^{\displaystyle \bf{*}}$};
\node [draw=none] at (512,372) {$\text{8}$};
\node [draw=none] at (65,378) {$\text{17}$};
\node [draw=none] at (174,393) {$\text{19}$};
\node [draw=none] at (288,412) {$\text{2}$};
\node [draw=none] at (392,439) {$\text{11}$};
\node [draw=none] at (493,468) {$\text{8}$};
\node [draw=none] at (59,474) {$\text{5}$};
\node [draw=none] at (167,490) {$\text{7}$};
\node [draw=none] at (272,507) {$\text{10}$};
\node [draw=none] at (378,527) {$\text{12}$};
\node [draw=none] at (489,551) {$\text{7}$};
\node [draw=none] at (41,555) {$\text{2}$};
\node [draw=none] at (146,568) {$\text{25}$};
\node [draw=none] at (266,598) {$\text{7}$};
\node [draw=none] at (366,617) {$\text{5}$};
\node [draw=none] at (470,642) {$\text{12}$};
\node [draw=none] at (29,652) {$\text{17}$};
\node [draw=none] at (139,662) {$\text{26}$};
\node [draw=none] at (244,685) {$\text{33}^{\displaystyle \bf{*}}$};
\node [draw=none] at (326,679) {$\text{23}$};
\node [] at (457,684) {$\text{14}$};
\addplot  [only marks, mark size=0.26cm, fill opacity=0.9, draw opacity=0.9, scatter, scatter src=explicit]
table [x=x, y=y, meta=colordata]{%
x                      y                      colordata
0.2865 6.51 0.312142070174261
1.394 6.673 0.685085927040432
2.431 6.855 0.914646248532035
3.563 7.085 0.510038617619258
4.217 7.206 0.268309670841333
0.420 5.595 0.0312786093990236
1.475 5.754 0.3733020503831
2.63 6.02 0.647700301868446
3.655 6.205 0.0956116641490765
4.703 6.406 0.292572508784498
0.55 4.76 0.192803655791449
1.645 4.92 0.190913094921226
2.725 5.095 0.224785003737408
3.783 5.315 0.111976513627961
4.805 5.575 0.219931620010259
0.664 3.788 0.431622108593821
1.738 3.90 0.396370445602672
2.91 4.155 0.147413625198577
3.915 4.387 0.506321106674979
4.941 4.6605 0.22862013464441
0.805 2.895 0.389295244783753
1.875 3.01 0.248893015403111
2.973 3.22 0.527589436383504
4.06 3.50 0.942273497490569
5.105 3.73 0.432355433055291
0.93 1.75 0.512572727714095
2.045 1.935 0.573277350638577
3.14 2.11 0.302103086335453
4.215 2.325 0.584277457601353
5.27 2.58 0.334918815894527
1.095 0.78 0.403516938521447
2.165 0.99 0.548737505681162
3.29 1.19 1.1184029105688
4.33 1.38 0.660896781604525
5.06 1.53 0.639444820729708
};

\end{axis}

\end{tikzpicture}
    \caption{The plot of the $35$ intersections in the northeast of San Francisco. We denote the number of crimes in each area by the number over each location. We denote the critical intersections by labeling time with a \textquotedblleft$*$\textquotedblright~next to the number of crimes.} 
    \label{fig:agent_distribution}
\end{figure}

Fig.~\ref{fig:agent_distribution} shows $35$ intersections in San Francisco, CA~\cite{wu2019reward}.
We use a factored MDP to describe the transition probabilities of the police officers between the intersections.
We also show the number of crimes that occurred in October 2018 around each intersection\footnote{The crime data can be found here https://bit.ly/35CDTJl.}.
A police officer obtains a reward equal to the number of crimes in an intersection if that officer monitors the intersection.
We incentivize the police officers to monitor intersections with higher crime rates with this reward function.
\looseness=-1%

There are some critical intersections on the map (see the intersections with \textquotedblleft$*$\textquotedblright in Fig.~\ref{fig:agent_distribution}).
We consider a GTL specification for the critical intersections to ensure that the critical states are monitored often by a police officer.
Specifically, for each critical intersection $s_{\textrm{crit}}$, we assign a police officer $v_i$ with the specification $\varphi = \Box_{[0, 20]}\left(\Diamond_{\leq 3} (s_{\textrm{crit}} \vee  \exists^{1} \bigcirc s_{\textrm{crit}})\right)$. 
The specification $\varphi$ reads as \textquotedblleft each critical intersection $s_{\textrm{crit}}$ should be monitored every three time steps by either the police officer $v_i$ or a neighboring police officer of $v_i$\textquotedblright for the next 20 time steps.\looseness=-1

We consider an example with 15 police officers, where each police officer is responsible for nine intersections on a $3 \times 3$ grid. 
For example, a police officer is responsible for intersections between $5\leq x_2 \leq 7$ and $3 \leq x_1 \leq 5$, and another police officer is responsible for intersections between $5\leq x_2 \leq 7$ and $2 \leq x_1 \leq 4$. 
The objective of the police officers is to maximize the expected reward by monitoring states with the highest crime rates while satisfying the GTL specification $\varphi$. 
\looseness=-1

For $\lambda=0.9$, Fig.~\ref{fig:agent_distribution} shows the resulting assignment for the police officers with $\beta=1$ after 500 iterations of Algorithm 1.
The computation time of the approach was 4.1 seconds. 
The results show that the GTL specification $\varphi$ is satisfied by the resulting assignment.
We also observe that the average number of police officers that monitor the intersections with higher crime rates (critical or non-critical) is higher to maximize the expected reward while satisfying the GTL specification $\varphi$.

\begin{figure}[t]
\centering
\begin{tikzpicture}
\begin{axis}[%
width=2.77in,
height=1.0in, 
at={(2.167in,0.898in)},
scale only axis,
xmin=1,
xmax=100,
    minor tick num=5,
ymajorticks=true,
yminorticks=true,
xmajorticks=true,
xminorticks=true,
xmajorgrids=true,
xminorgrids=true,
ymajorgrids=true,
yminorgrids=true,
    grid=both,
    grid style={line width=.1pt, draw=gray!10},
    major grid style={line width=.2pt,draw=gray!50},
legend style={at={(1.0,0.05)},anchor=south east,font=\fontsize{8.5}{8.5}\selectfont},
xlabel style={font=\color{white!0!black}},
xlabel={Number of Intersections for each Police Officer},
ymin=0.1,
ymax=3600,
ymode=log,
ylabel style={font=\color{white!0!black}},
ylabel={Running Time (s)},
axis background/.style={fill=white},
title style={font=\bfseries},
title={},
ytick={1,10,100,1000,3600},
    yticklabels={$10^0$,$10^1$,$10^2$,$10^3$,TO},
]
\addplot table[x=x,y=y] {data_police.dat}; 
\addlegendentry{Local Policies}
\addplot table[x=x,y=y] {data_police2.dat}; 
\addlegendentry{Neighboring Policies}

\addplot [name path=upper1,draw=none] table[x=x,y expr=\thisrow{y}+10*\thisrow{ey}] {data_police.dat};
\addplot [name path=lower1,draw=none] table[x=x,y expr=\thisrow{y}-10*\thisrow{ey}] {data_police.dat};
\addplot [fill=blue!10] fill between[of=upper1 and lower1];
\addplot [name path=upper,draw=none] table[x=x,y expr=\thisrow{y}+10*\thisrow{ey}] {data_police2.dat};
\addplot [name path=lower,draw=none] table[x=x,y expr=\thisrow{y}-10*\thisrow{ey}] {data_police2.dat};
\addplot [fill=red!10] fill between[of=upper and lower];
\end{axis}
\end{tikzpicture}
\caption{The running times of the police officer example with 15 agents and varying number of intersections. 
The method with local policies runs two orders of magnitude faster and can solve problems with larger state spaces before the time-out.\looseness=-1}
\label{fig:num_agent_police}
\end{figure}
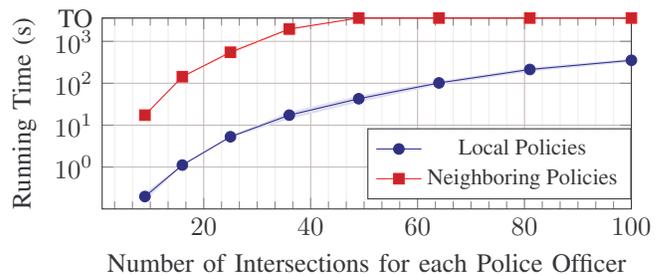

\noindent\emph{Comparison between proposed methods:} We now demonstrate the comparison between the two proposed methods on larger state spaces for each agent. 
We consider an example with 15 police officers, but now, each police officer is responsible for a varying number of intersections. 
We show the running time of both of the methods in Fig.~\ref{fig:num_agent_police} with a varying number of intersections for each police officer.
The shades indicate the standard variation of the running times over 20 trials.
We set the time-out (labeled as \textquotedblleft TO\textquotedblright) to one hour.

The results in Fig.~\ref{fig:num_agent_police} show that the synthesis method with local policies for each agent scales much better and able to solve tasks with much larger state spaces for each agent compared to computing neighboring policies.
The synthesis method with local policies can solve an urban security task with $100$ intersections faster than our previous method with $25$ intersections.
Additionally, this method is, on average, two orders of magnitude faster.
The mean difference between the obtained expected reward is $1.13\%$ between the two methods, where the method with neighboring policies incurs higher expected rewards.

\section{Appendix B}
In this appendix, we provide proofs for all results presented
in this paper.

\textbf{Proof of Theorem 1:} We utilize the results of~\cite{DBLP:journals/lmcs/EtessamiKVY08} to relate the variables $o_i(\hat{s}_{N(i)},\hat{\act}_{N(j)})$ with the expected residence time in states. In \cite{DBLP:journals/lmcs/EtessamiKVY08}, it is shown that for an MDP $\mathcal{M}$, the variables 
$o(s,\act)$ satisfy the constraint in~\eqref{eq:policylp:probthresh} and corresponds to the expected time of taking an action $\act$ state-action pair $(s,\act)$ in an induced MC $\mathcal{M}^{\pi}$. Additionally, the variable  $o(s)$ for an accepting  state $s$ corresponds to the reachability probability of states $s\in Acc$.

Then, for states $s\in S\setminus Acc$,%
\begin{align}
\label{state_residence}
\displaystyle o^{\sched}(s)=\sum_{\alpha\in \Act}o(s,\act).
\end{align}%
Additionally, if $\sum_{\act\in \Act}o(s,\act)>0$, we can compute the policy by~\eqref{eq:occupmeasure}. For each agent $v_i$, we can define a MDP with a state space $S_{N(i)}$ and an action space $\Act_{N(i)}$. Then, the constraints~\eqref{eq:graph_mdp_flow} and~\eqref{eq:graph_mdp_flow2} ensure that the variable $o_{i}(\hat{s}_{N(i)},\hat{\act}_{N(i)})$ for each state $\hat{s}_{N(i)} \in \bar{S}^{\mathrm{p}}_{N(i)}$ and action $\hat{\act}_{N(i)}\in\Act_{N(i)}$ defines the occupancy measure of a state-action pair for $\sched_i$. By adding the constraints in~\eqref{eq:graph_mdp_neighbor} we ensure that the time spent in the set of states $\hat{s}_{N(i,j)}$ and taking the set of actions $\hat{\act}_{N(i,j)}$ is equal for the policies of agents $v_i$ and $v_j$, and therefore the policies of the agents are consistent. Therefore, we conclude the policy $\sched$ is optimal for each agent $v_i$ for $\mathcal{M}_{\textrm{p}}$.
\balance

\added{\textbf{Proof of Lemma 1:}
 For each pair of agents $v_i, v_j \in V$, we assume that the set $N(i,j)$ is nonempty. 
 Otherwise, we do not add the constraints in~\eqref{eq:graph_mdp_local_neighbor_occup} for the agents $v_i$ and $v_j$.
 Given local policies for agents in $N(i,j)$, $N(i\setminus j)$, and $N(j\setminus i)$, we construct the occupancy measures for each state $\hat{s}_{N(i,j)}=\lbrace s_{k_1},\ldots,s_{k_{N(i,j)}}\rbrace$ in $S_{N(i,j)}$ and $t \in \bar{T}_\varphi$ by
 \begin{align}
     &\displaystyle \sum_{\hat{s}_{N(i/j)}\in S_{N(i,j)}}o_i(\hat{s}_{N(i,j)},\hat{s}_{N(i/j)},t)=\\
    &\displaystyle \sum_{\hat{s}_{N(j/i)}\in S_{N(i,j)}}o_j(\hat{s}_{N(i,j)},\hat{s}_{N(j/i)},t)=\\
     &\prod_{s_k \in \hat{s}_{N(i,j)}}\dfrac{o_k(s_k,t)}{\sum_{s_k' \in S_k}o_k(s_k',t)}=\prod_{s_k \in \hat{s}_{N(i,j)}}o_k(s_k,t)~\label{proof:eq}.
 \end{align}
The equality in~\eqref{proof:eq} holds as the behavior of each agent is independent given local policies for each agent in $N(i,j)$ and $\sum_{s_k \in S_k}o_k(s_k,t)=1$ by the definition of $\mu$.}

\begin{IEEEbiography}[{\includegraphics[width=1in,height=1.25in,clip,keepaspectratio]{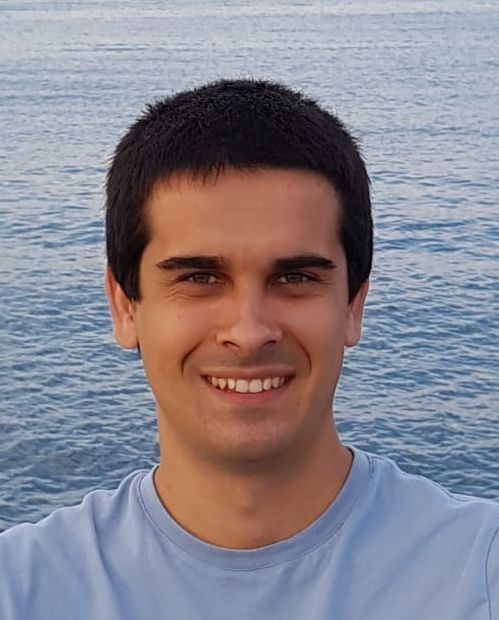}}]{Murat Cubuktepe} joined the Department of Aerospace Engineering at the University of Texas at Austin as a Ph.D. student in Fall 2015. He received his B.S degree in Mechanical Engineering from Bogazici University in 2015 and his M.S degree in Aerospace Engineering and Engineering Mechanics from the University of Texas at Austin in 2017. His research focuses on developing theory and algorithms for verified learning and control for autonomous systems.
\end{IEEEbiography}

\begin{IEEEbiography}[{\includegraphics[width=1in,height=1.25in,clip,keepaspectratio]{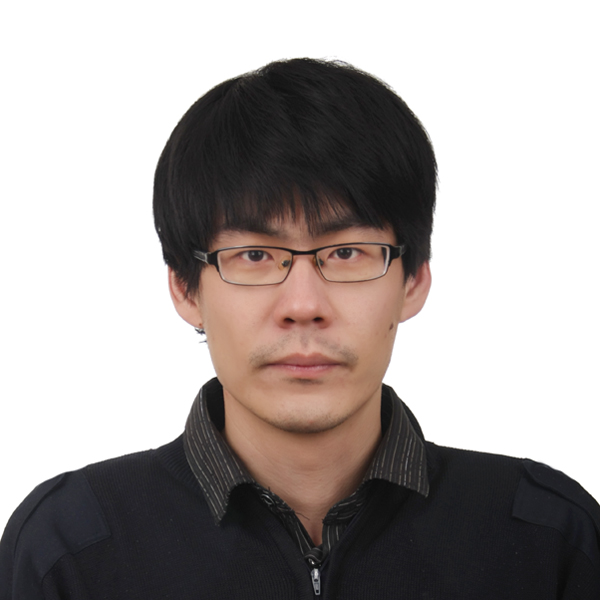}}]{Zhe Xu}
received the B.S. and M.S. degrees in Electrical Engineering from Tianjin University, Tianjin, China, in 2011 and 2014, respectively. He received the Ph.D. degree in Electrical Engineering at Rensselaer Polytechnic Institute, Troy, NY, in 2018. He is currently an assistant professor in the School for Engineering of Matter, Transport, and Energy at Arizona State University. Before joining ASU, he was a postdoctoral researcher in the Oden Institute for Computational Engineering and Sciences at the University of Texas at Austin, Austin, TX. His research interests include formal methods, autonomous systems, control systems and reinforcement learning. 

\end{IEEEbiography}

\begin{IEEEbiography}[{\includegraphics[width=1in,height=1.25in,clip,keepaspectratio]{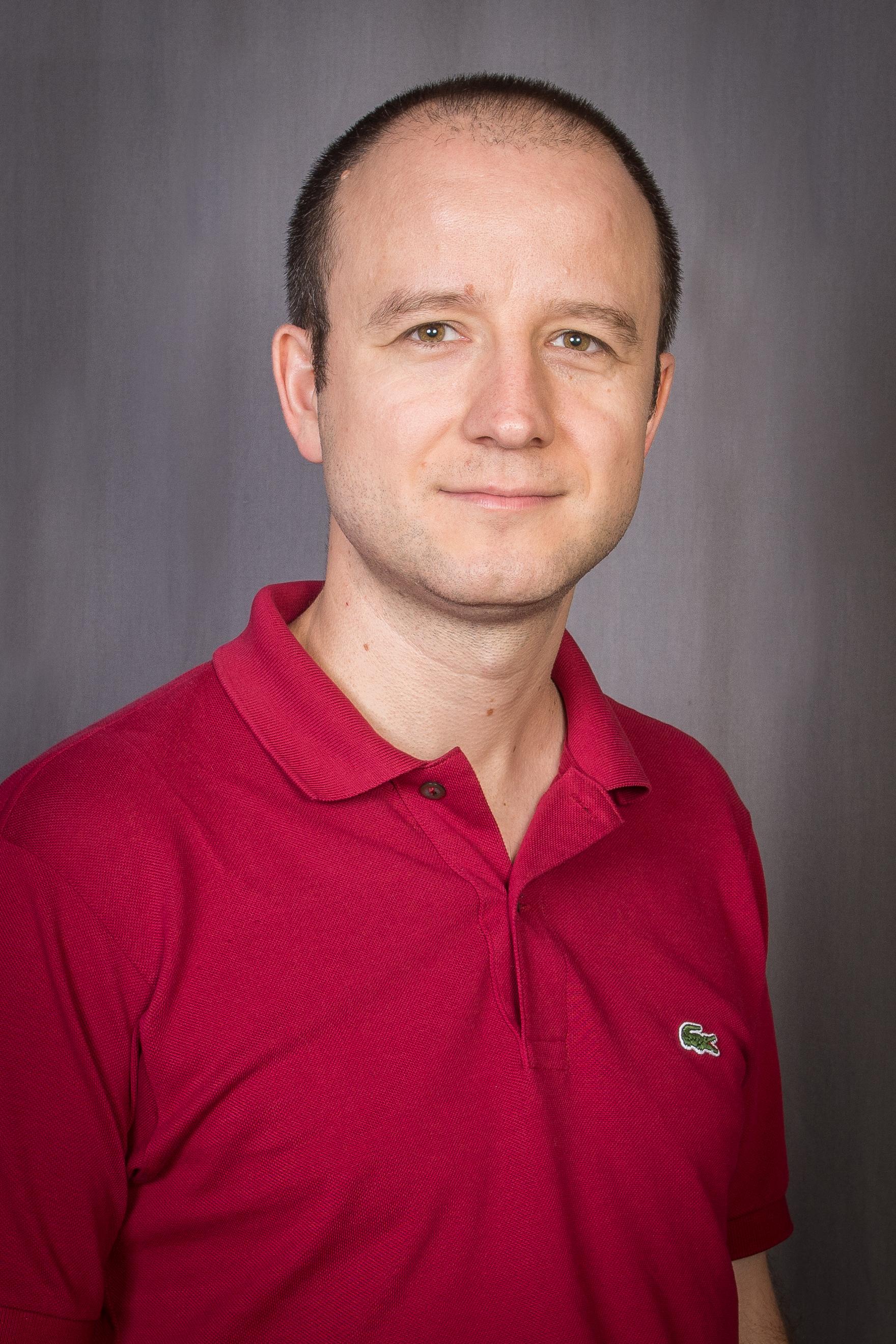}}]{Ufuk Topcu} is an associate professor in the Department of Aerospace Engineering and Engineering Mechanics and the Oden Institute at The University of Texas at Austin. He received his Ph.D. degree from the University of California at Berkeley in 2008. His research focuses on the theoretical, algorithmic, and computational aspects of design and verification of autonomous systems through novel connections between formal methods, learning theory and controls.
\end{IEEEbiography}

\end{document}